\documentclass[aps,prb,twocolumn,reprint,superscriptaddress,citeautoscript]{revtex4-2}

\usepackage{physics}
\usepackage{graphicx}
\usepackage{amsmath,amssymb}
\usepackage[normalem]{ulem}
\usepackage{bm}
\usepackage{upgreek}

\usepackage{xcolor}

\usepackage{hyperref}
\hypersetup{
	citecolor = blue,
	colorlinks = true,
	urlcolor = blue
}

\begin{document}
	
\title{Spin pumping into quantum spin nematic states}

\author{Takuto Ishikawa}
\affiliation{Institute for Solid State Physics, University of Tokyo, Kashiwa 277-8581, Japan}
\author{Wolfgang Belzig}
\affiliation{Department of Physics, University of Konstanz, 78457 Konstanz, Germany}
\author{Takeo Kato}
\affiliation{Institute for Solid State Physics, University of Tokyo, Kashiwa 277-8581, Japan}
\email{kato@issp.u-tokyo.ac.jp}

\begin{abstract}
We theoretically study spin pumping into a spin-nematic state in a junction system composed of a ferromagnetic insulator and a spin-nematic insulator, described by using the spin-1 bilinear-biquadratic model.
We analyze the increase of the Gilbert damping in ferromagnetic resonance (FMR) due to an interfacial exchange coupling within a mean-field theory based on the Schwinger boson method. We find that the two Schwinger bosons contribute in distinct ways to spin pumping. We report a detailed dependence of the spin pumping on a resonant frequency and a magnetic field, along with a comparison to the case of the canted antiferromagnetic insulator. Our findings establish spin pumping as a relevant technique to verify and characterize spin-nematic phases.
\end{abstract}

\maketitle

\section{Introduction} 

In the development of spintronics, the technique of spin pumping driven by ferromagnetic resonance (FMR) has emerged as a versatile tool for injecting spins into materials adjacent to ferromagnets~\cite{Tserkovnyak2002,Simanek2003,Zutic2004,Tserkovnyak2005,Hellman2017,Tsymbal2019}.
Spin pumping can also serve as a probe for spin excitations of target materials through the Gilbert damping, which can be extracted from the width of the FMR peak.
In fact, enhancement of the Gilbert damping due to interfacial coupling with target materials has been shown to contain valuable information regarding their spin excitations~\cite{Kamra2016,Kamra2017,Han2020,Yang2018,Rozsa2018,Qiu2016,Yamamoto2021,Ominato2020a,Ominato2020b,Yama2021,Inoue2017,Silaev2020,Silaev2020b,Simensen2021,Fyhn2021,Ominato2022a,Ominato2022b,Funato2022,Sun2023,Funaki2023,Sun2023b,Yama2023a,Fukuzawa2023,Haddad2023,Furuya2024,Yama2024proceeding}.

Spin pumping into exotic materials with hidden orders represents a significant challenge in identifying novel states of matter and understanding their quantum nature. 
Among these exotic hidden orders, the spin-nematic state, characterized by spin-rotation symmetry breaking without any accompanying magnetic order, stands as a representative example.
The spin-nematic state is described by quadrupolar order or higher multipolar order and has been extensively studied theoretically, using models such as the bilinear-biquadratic exchange model~\cite{Blume1969,Chen1971,Andreev1984,Papanicolaou1988,Tanaka2001,Bhattacharjee2006,Lauchli2006,Tsunetsugu2006,Li2007,Tsunetsugu2007}, the $J_1$-$J_2$ frustrated model~\cite{Chubukov1991,Shannon2006,Ueda2007,Vekua2007,Hikihara2008,Sato2009,Sudan2009,Zhitomirsky2010,Sato2011,Shindou2011,Momoi2012,Sato2013,Ueda2013,Jiang2023}, and the dimer model~\cite{Momoi2000,Wang2018,Yokohama2018,Hikihara2019}.
These theoretical models predict the existence of ferro- or antiferro-quadrupolar states.
The realization of the spin nematic state has also been discussed in a variety of materials such as ${\mathrm{LiCuVO}}_{4}$~\cite{Buttgen2014,Orlova2017,Yoshida2017} and ${\mathrm{BaCdVO}}(\mathrm{PO}4)_{2}$~\cite{Nath2008,Skoulatos2019,Povarov2019}. In the absence of dipole moments, the unique properties of the SN state are difficult to observe using static magnetization measurements. Conventional approaches employ NMR relaxation rates in an attempt to identify SN by focusing on a feature appearing in the low-temperature behavior. However, this feature is challenging to detect experimentally, due to the difficulty in separating the two magnetic excitations~\cite{Momoi2024}.
Recently, a study of the spin Seebeck effect in the spin-nematic state has also been conducted~\cite{Hirobe2019}.

\begin{figure}[tb]
\begin{center}
\includegraphics[clip,width=8.0cm]{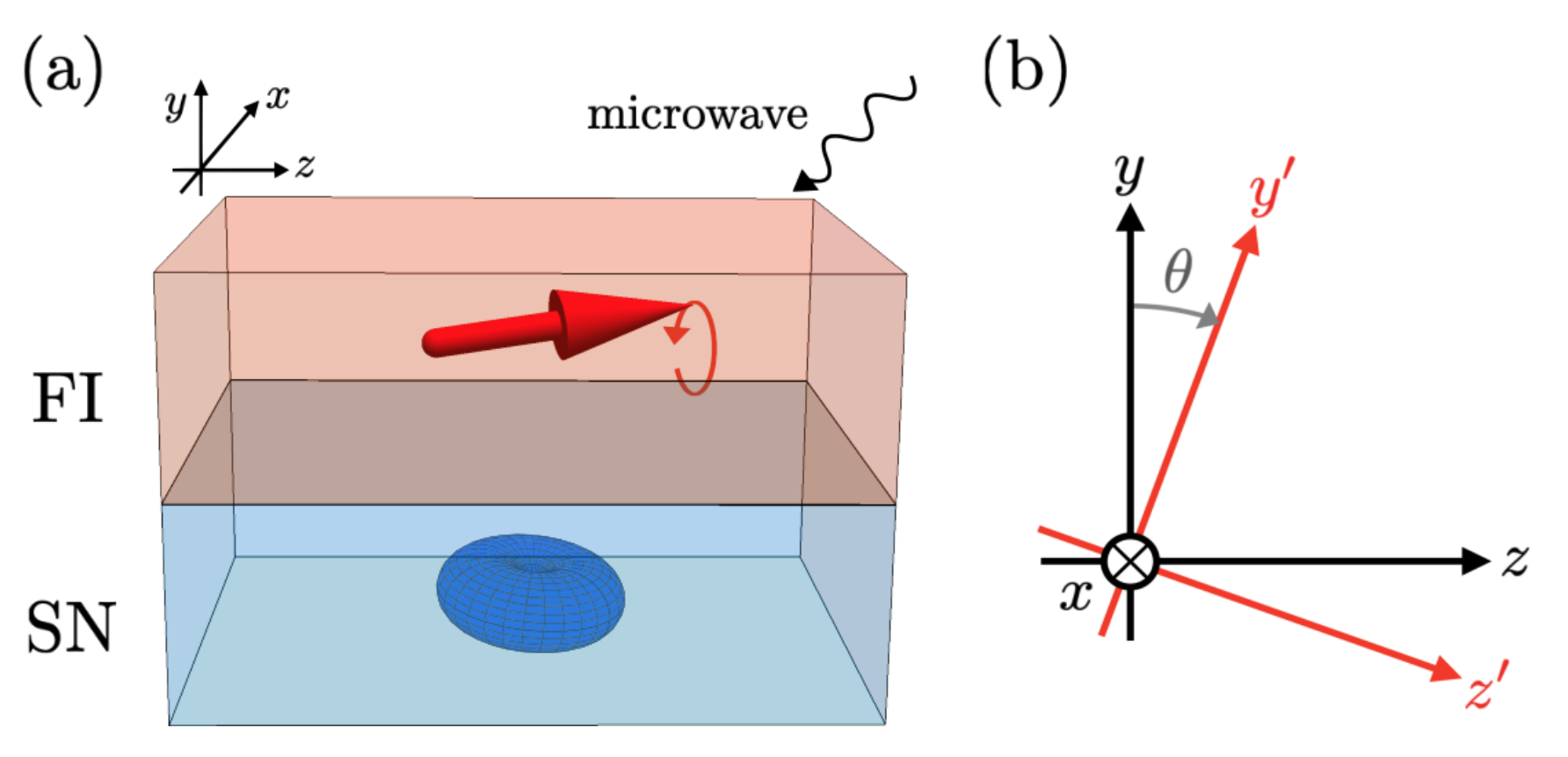}
\caption{(a) Magnetic junction composed of a ferromagnetic insulator (FI) and a spin nematic (SN) insulator. (b) Coordinate transformation between the laboratory coordinates ($x,y,z$) and the magnetization-fixed coordinates ($x',y',z'$).}
\label{fig:setup}
\end{center}
\end{figure}

In this work, we theoretically calculate the enhancement of the Gilbert damping resulting from spin pumping into a spin-nematic insulator. 
We consider a magnetic junction composed of a spin-nematic (SN) insulator and a ferromagnetic insulator (FI) subjected to microwave irradiation (see Fig.~\ref{fig:setup}). 
To discuss qualitative features of spin pumping, we employ a simple model based on the bilinear-biquadratic exchange model to realize the spin-nematic state.
We derive an analytical expression for the enhancement of the Gilbert damping by the Schwinger-boson mean-field method and second-order perturbation theory with respect to interface coupling.
Furthermore, we examine how the enhancement of Gilbert damping depends on the FMR frequency and magnetic field.
For comparison with other magnetic states, we also calculate the enhancement of the Gilbert damping resulting from spin pumping into a canted antiferromagnetic insulator, and discuss the difference to the spin nematic.

The rest of this work is organized as follows.
In Sec.~\ref{sec:model}, we introduce our theoretical model for junction systems and explain the Schwinger-boson mean-field theory for the spin-nematic phase. 
In Sec.~\ref{sec:formulation}, we theoretically calculate the enhancement of the Gilbert damping due to interfacial exchange coupling within second-order perturbation.
In Sec.~\ref{sec:Result}, we show how the increase of the Gilbert damping depends on the orientation and strength of the external magnetic field and compare its results with that of a canted antiferromagnetic insulator.
Finally, we summarize our results in Sec.~\ref{sec:Summary}. Appendix~\ref{app:spin_sus_nematic} and appendix~\ref{app:spin_sus_cant} are devoted to a detailed derivation of analytic expressions.

\section{Model} 
\label{sec:model}

We consider a junction system composed of a ferromagnetic insulator (FI) and a spin-nematic insulator (SN).
The model Hamiltonian is expressed as
\begin{align}
\mathcal{H}_{{\rm FI} / {\rm SN}}&=\mathcal{H}_{\rm FI}+\mathcal{H}_{\rm SN}+\mathcal{H}_{\rm int},
\end{align}
where $\mathcal{H}_{\rm FI}$, $\mathcal{H}_{\rm SN}$, and $\mathcal{H}_{\rm int}$ represent FI, SN, and an interfacial exchange coupling, respectively.
Explicit forms of each Hamiltonian are provided in the following subsections.

To identify the feature of the nematic phase through the FMR experiment, we also consider a junction composed of FI and a canted antiferromagnetic insulator (CAF).
We note that the SN and CAF phases have a few common features; (i) they have two excitation modes and (ii) they do not have the U(1) symmetry, leading to the breakdown of the conservation law of a magnon number (or a spin component in a specific direction).
The model Hamiltonian of the FI/CAF junction is expressed as
\begin{align}
\mathcal{H_{{\rm FI} / {\rm CAF}}}&=\mathcal{H}_{\rm FI}+\mathcal{H}_{\rm CAF}+\mathcal{H}_{\rm int},
\end{align}
where $H_{\rm cant}$ describes CAF and $\mathcal{H}_{\rm int}$ represents an interfacial exchange coupling with the same form as the FI/SN junction system.
By comparing the results for the SN/FI and CAF/FI junctions, we clarify how to identify the feature of the SN phase in the FMR experiment.

\subsection{Ferromagnetic insulator (FI)}

To describe the FI, we consider the Heisenberg model with axial anisotropy, whose Hamiltonian is expressed as.
\begin{align}
\mathcal{H}_{\rm FI} = J \sum_{\langle i,j \rangle}
{\cal \bm{S}}_{i} \cdot {\cal \bm{S}}_{j}-\lambda_L \sum_{i} ({\cal S}^z_{i})^2,
\end{align}
where ${\cal \bm{S}}_{i}$ is a localized spin with amplitude of $S_0$, $J$ ($<0$) is a ferromagnetic exchange coupling, $\langle i,j \rangle$ indicates a pair of neighboring sites, and $\lambda_L$ denotes the strength of the axial anisotropy.
For simplicity, we assume that the anisotropy energy proportional to $\lambda_L$ is much larger than the Zeeman energy induced by an external magnetic field.
Consequently, the spin polarization in the FI is aligned along the $z$-axis throughout this paper.
Using the Holstein-Primakoff transformation and the independent magnon approximation, the Hamiltonian of the FI is rewritten in the long-wavelength approximation as $\mathcal{H}_{\rm FI} = \sum_{\bm k}\hbar \omega_{\bm k} b_{\bm k}^\dag b_{\bm k}$, where $b_{\bm k}$ ($b_{\bm k}^\dagger$) is the annihilation (creation) operator of magnons with wavenumber ${\bm k}$, $\omega_{\bm k} = 2S_0\lambda_L + {\cal D}k^2$ is a magnon dispersion, and ${\cal D}$ is a spin stiffness.
Since the external microwave induces a uniform spin precession, it is sufficient to consider the Hamiltonian for ${\bm k}={\bm 0}$:
\begin{align}
{\cal H}_{\rm FI} = \hbar \omega_{\bm 0} b_{\bm 0}^\dagger b_{\bm 0}, 
\end{align}
where $\omega_{\bm 0} = 2S_0\lambda_L$.

\subsection{Spin-nematic (SN) state}

As a simple model to realize the spin-nematic state, we consider a spin-1 bilinear-biquadratic model on a triangular lattice, whose Hamiltonian is given as
\begin{align}
\mathcal{H}_{\rm SN} &= \sum_{\left<i,j\right>}
\left[
J_1 \, {\bm{S}}_{i} \cdot {\bm{S}}_{j}
+J_2 \, ({\bm{S}}_{i} \cdot {\bm{S}}_{j})^{2}
\right] \nonumber \\
&\hspace{5mm}-h \sum_{i} S^{z'}_{i} -D \sum_{i} (S^{x'}_{i})^2,
\end{align}
where ${\bm{S}}_{i}$ denotes a localized spin, $J_1$ and $J_2$ denote the strengths of the bilinear and biquadratic couplings, respectively, and $h$ represents an external magnetic field.
For simplicity in the calculation, we assume $D$ is much smaller than $J_1$, $J_2$, and $h$, existing solely to determine the direction of quadrupolar ordering. We employ a triangular lattice following a previous study~\cite{Lauchli2006}
though spin nematic states can appear also in a square lattice.
We introduced a new coordinate $(x',y',z')$, to the original ordinate as
\begin{align}
\begin{pmatrix}x \\y \\z 
\end{pmatrix}
= 
\begin{pmatrix}
1 & 0 & 0 \\
0 & \cos \theta & -\sin \theta \\
0 & \sin \theta & \cos \theta
\end{pmatrix}
\begin{pmatrix}x' \\y' \\z' 
\end{pmatrix},
\label{eq:CoordinateTransformation}
\end{align}
and the external magnetic field is applied along the $z'$ direction (see Fig.~\ref{fig:setup}(b)).
This bilinear-biquadratic model is known to exhibit ferromagnetic quadrupole ordering (spin-nematic phase) within a finite range of the ratio $J_1/J_2$~\cite{Lacroix2011}.
To construct a mean-field theory for the spin-nematic phase, we introduce the operators of the SU(3) Schwinger boson as follows~\cite{Lauchli2006,Li2007}:
\begin{align}
S^{x'}_i&=-i[a_{y'}^\dag(i) a_{z'}(i) - a_{z'}^\dag(i) a_{y'}(i)], \label{SxdashBoson} \\
S^{y'}_i&=-i[a_{z'}^\dag(i) a_{x'}(i) - a_{x'}^\dag(i) a_{z'}(i)], 
\label{SydashBoson} \\
S^{z'}_i&=-i[a_{x'}^\dag(i) a_{y'}(i) - a_{y'}^\dag(i) a_{x'}(i)].
\label{SzdashBoson} 
\end{align}
In the absence of an external magnetic field, a mean-field theory for the spin nematic state is constructed by replacing one of the Schwinger bosons with its mean value.
The direction of the ferro-quadrupolar state, i.e., the component of the condensed Schwinger bosons, is perpendicular to the magnetic field. 
In our calculation, we assume that condensation occurs for the $y'$-component boson denoted as
\begin{align}
a_{y'}(i), a_{y'}^\dagger(i) \rightarrow \sqrt{M-a_{x'}^\dagger(i)a_{x'}(i)-a_{z'}^\dagger(i)a_{z'}(i)},
\end{align}
where $M$ represents the saturation amplitude of the quadrupole ordering.
Although $M$ equals the magnitude of the localized spin, i.e. $M=1$ for the spin-1 system, we treat $M$ as a control parameter.

In the presence of an external magnetic field in the $z'$ direction, the quadrupole ordering loses the axial symmetry around the $y'$ axis.
Therefore, we need to prepare new Schwinger bosons as
\begin{align}
\begin{pmatrix}
a_{x'}^\dag \\
a_{y'}^\dag \\
a_{z'}^\dag 
\end{pmatrix}
=
\begin{pmatrix}
\cos\frac{\mu}{2} & i\sin\frac{\mu}{2} & 0 \\
i\sin\frac{\mu}{2} & \cos\frac{\mu}{2} & 0 \\
0 & 0 & 1 
\end{pmatrix}
\begin{pmatrix}
a_{\perp}^\dag \\
a_{\parallel}^\dag \\
a_{z'}^\dag 
\end{pmatrix},
\label{bosontrans}
\end{align}
with a finite magnetic moment $m=M\sin\mu$,
and then assume condensation of the boson operator $a_\parallel$ as
\begin{align}
a_{\parallel}(i), a_{\parallel}^\dagger(i) \rightarrow \sqrt{M-a_{\perp}^\dagger(i)a_{\perp}(i)-a_{z'}^\dagger(i)a_{z'}(i)}.
\label{condensepara}
\end{align}
The magnetization is proportional to the magnetic field from the mean field result,
\begin{align}
m =\frac{h}{z(J_1-J_2)}
\end{align}
where $z$ ($=6$) is the number of nearest neighbor sites.
By expanding the Hamiltonian with respect to $a_{z'}$, $a_{\perp}$, and their Hermite conjugates and by leaving the leading contribution with respect to $1/M$, the Hamiltonian is approximately rewritten as
\begin{align}
{\cal H}_{\rm SN} &= \frac{zM}{2} \sum_{\nu = \perp, z'} \sum_{\bm k} \Bigl\{ 2A_{{\bm k}\nu} a_{{\bm k}\nu}^\dagger a_{{\bm k}\nu} \nonumber \\
& \hspace{5mm} + B_{{\bm k}\nu}\bigl[a_{{\bm k}\nu}a_{-{\bm k}\nu} + a^\dagger_{{\bm k}\nu} a^\dagger_{-{\bm k}\nu}\bigr] \Bigr\},
\end{align}
where the coefficients are given as
\begin{align}
&A_{{\bm k}{z'}} = J_1 \gamma_{\bm k}-J_2, \\
&B_{{\bm k}{z'}} = (J_2-J_1) \cos \mu \, \gamma_{\bm k}, \\
&A_{{\bm k}\perp} = (J_1 \cos^2 \mu + J_2 \sin^2 \mu ) \gamma_{\bm k}-J_2, \\
&B_{{\bm k}\perp} = (J_2-J_1) \cos^2 \mu \, \gamma_{\bm k}, \\
& \gamma_{\bm k} = \frac 1 z  \sum_{\delta} e^{i {\bm k} \cdot {\bm r}_\delta},
\end{align}
where ${\bm r}_\delta$ denotes a displacement vector spanning a site to its neighboring site $\delta$. Note the formal similarity of the terms with $B_{{\bm k} \nu}$  to quantum squeezing known in quantum optics which has been pointed out by ~\cite{Kamra2016}.
This Hamiltonian can be diagonalized by the Bogoliubov transformation,
\begin{align}
a_{{\bm k}\nu}^\dag &= \cosh \xi_{{\bm k}\nu} \, \alpha_{{\bm k}\nu}^\dag +
\sinh \xi_{{\bm k}\nu} \, \alpha_{-{\bm k}\nu}, \label{Bogoliubov1} \\
a_{{\bm k}\nu} &= \cosh \xi_{{\bm k}\nu} \, \alpha_{{\bm k}\nu} +
\sinh \xi_{{\bm k}\nu} \, \alpha_{-{\bm k}\nu}^\dag ,\label{Bogoliubov2}
\end{align}
where $\xi_{{\bm k}\nu}$ is chosen to satisfy
\begin{align}
\tanh (2\xi_{{\bm k}\nu}) = -\frac{B_{{\bm k}\nu}}{A_{{\bm k}\nu}}.
\end{align}
By dropping a constant term, we finally obtain the following Hamiltonian:
\begin{align}
\mathcal{H}_{\rm SN}&=\sum_{\nu=z',{\perp}} \sum_{\bm k} 
\varepsilon_{{\bm k}\nu} \alpha_{{\bm k}\nu}^\dag \alpha_{{\bm k}\nu},
\label{HSNdiag}
\end{align}
where $\varepsilon_{{\bm k}\nu}$ is a dispersion of the Schwinger bosons given as
\begin{align}
\varepsilon_{{\bm k}\nu} = \frac{zM}{2}\sqrt{A_{{\bm k}\nu}^2-B_{{\bm k}\nu}^2}, \quad (\nu = z',\perp).
\end{align}

\subsection{Canted antiferromagnetic insulator (CAF)}
\label{subsec:canted}

\begin{figure}[tb]
\begin{center}
\includegraphics[clip,width=8.0cm]{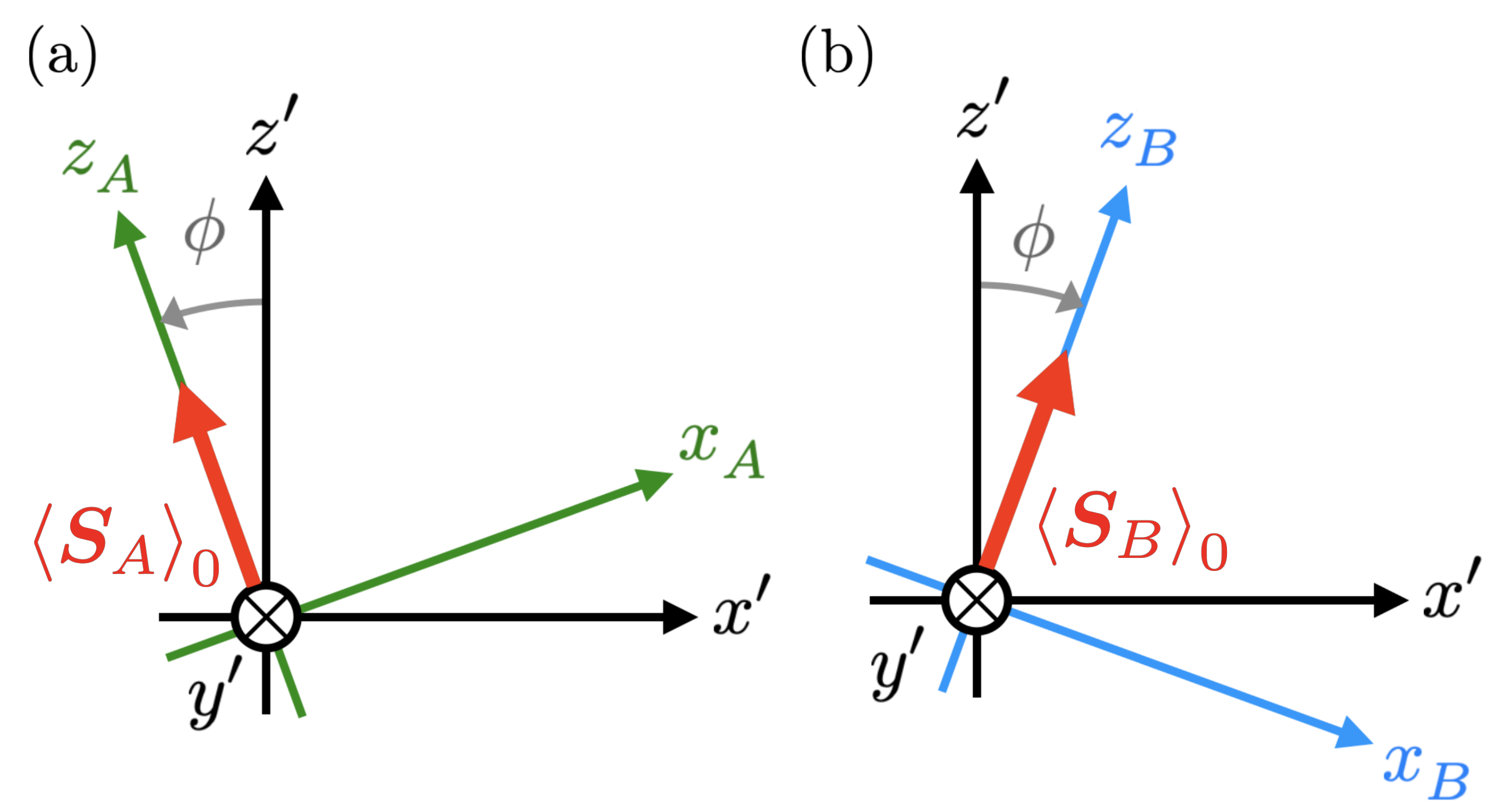}
\caption{Coordinate transformation between the coordinates $O$-$x'y'z'$ and the canted coordinates for each sublattices (a) $O$-$x_{\rm A} y_{\rm A} z_{\rm A}$ and (b) $O$-$x_{\rm B} y_{\rm B} z_{\rm B}$.}
\label{fig:canted_setup}
\end{center}
\end{figure}

To describe the CAF insulator, we consider the three-dimensional Heisenberg model on a cubic lattice, whose Hamiltonian is expressed as
\begin{align}
\mathcal{H}_{\rm CAF} = \sum_{\left<i,j\right>}
J_3 \, {\bm{S}}_{i} \cdot {\bm{S}}_{j}
-h \sum_{i} S^{z'}_{i} -D \sum_{i} (S^{x'}_{i})^2,
\end{align}
where ${\bm{S}}_{i}$ denotes a localized spin, $J_3$ ($>0$) is an antiferromagnetic exchange coupling, $h$ represents an external magnetic field, and the coordinate $(x',y',z')$ is defined by Eq.~(\ref{eq:CoordinateTransformation}) (see also Fig.~\ref{fig:setup}(b)).
We divide the lattice into A and B sublattices and further assume that $D$ is much smaller than $J_3$ and $h$, existing solely to determine the direction of magnetic ordering.
The spin at each site points to the $+z'$ direction for a strong magnetic field while it points to the $\pm x'$ direction for a weak magnetic field due to the small anisotropy $D$, depending on the sublattice.
To interpolate the rotation of the ordered spin as a function of the magnetic field, we introduce an angle $\phi$ and new coordinates, $(x_{\rm A},y_{\rm A},z_{\rm A})$ and $(x_{\rm B},y_{\rm B},z_{\rm B})$ as shown in Fig.~\ref{fig:canted_setup}(a) and (b):
\begin{align}
\begin{pmatrix}x' \\y' \\z' 
\end{pmatrix}
= 
\begin{pmatrix}
\cos \phi & 0 & -\sin \phi \\
0 & 1 & 0 \\
\sin \phi & 0 & \cos \phi
\end{pmatrix}
\begin{pmatrix}x_{\rm A} \\y_{\rm A} \\z_{\rm A} 
\end{pmatrix}, \label{eq:coordinate_caf1} \\
\begin{pmatrix}x' \\y' \\z'
\end{pmatrix}
= 
\begin{pmatrix}
\cos \phi & 0 & \sin \phi \\
0 & 1 & 0 \\
-\sin \phi & 0 & \cos \phi
\end{pmatrix}
\begin{pmatrix}x_{\rm B} \\ y_{\rm B} \\ z_{\rm B} 
\end{pmatrix}. \label{eq:coordinate_caf2}
\end{align}
In these coordinates, the averaged spin is represented as $(S_{\rm A}^x, S_{\rm A}^y, S_{\rm A}^z) = (0,0,S)$ and $(S_{\rm B}^x, S_{\rm B}^y, S_{\rm B}^z) = (0,0,S)$, where $S$ is an amplitude of the localized spin ${\bm S}_i$.
Using the Holstein-Primakoff transformation, we introduce magnons for each sublattice as follows,
\begin{align}
S^{+}_{{\rm A},i}&=\sqrt{2S} a_i , \ S^{-}_{{\rm A},i}=\sqrt{2S} a^\dag_i , \ S^{z}_{{\rm A},i}=S- a^\dag_i a_i \label{eq:sa_cant} \\
S^{+}_{{\rm B},i}&=\sqrt{2S} b_i , \ S^{-}_{{\rm B},i}=\sqrt{2S} b^\dag_i , \ S^{z}_{{\rm B},i}=S- b^\dag_i b_i .\label{eq:sb_cant}
\end{align}
With the Fourier transformations of the magnon operators, $a_{\bm k}$ and $b_{\bm k}$, we introduce new magnon operators as follows:
\begin{align}
\begin{pmatrix}a_{\bm k} \\ b_{-\bm k} 
\end{pmatrix}
= \frac{1}{\sqrt{2}}
\begin{pmatrix}
1 & 1 \\
-1 & 1 
\end{pmatrix}
\begin{pmatrix}c_{\bm k} \\ d_{-\bm k} 
\end{pmatrix}.\label{eq:bogoliubov_cant1}
\end{align} 
Leaving the leading-order terms of $1/S$, the Hamiltonian is approximately rewritten as
\begin{align}
{\cal H}_{\rm CAF} &= \sum_{\bm k} \Bigl\{
A_{\bm k} c_{{\bm k}}^\dagger c_{{\bm k}}-
\frac{C_{\bm k}}{2}(c_{{\bm k}} c_{{-\bm k}}+c_{{\bm k}}^\dagger c^\dag_{{-\bm k}})
\nonumber \\
& \hspace{5mm} +B_{\bm k} d_{{\bm k}}^\dagger d_{{\bm k}}+
\frac{C_{\bm k}}{2}(d_{{\bm k}} d_{{-\bm k}}+d_{{\bm k}}^\dagger d^\dag_{{-\bm k}}) \Bigr\}.
\end{align}
Here, the coefficients are given as
\begin{align}
A_{\bm k} &= 6J_3 S(1-2\cos^2\phi)+h\cos\phi \nonumber \\
& \hspace{5mm} +2J_3S\gamma_{\bm k}(1-\cos^2\phi),\\
B_{\bm k} &= 6J_3 S(1-2\cos^2\phi)+h\cos\phi \nonumber \\
& \hspace{5mm} -2J_3S\gamma_{\bm k}(1-\cos^2\phi),\\
C_{\bm k} &=2J_3 S\gamma_{\bm k} \cos^2\phi,
\end{align}
where $\gamma_{\bm k} =\cos k_x+\cos k_y+\cos k_z$. As for the SN state the terms with $C_{\bm k}$ lead to squeezing of the magnons.
To diagonalize the Hamiltonian, we consider the following Bogoliubov transformation
\begin{align}
\begin{pmatrix}c_{\bm k} \\ c^{\dag}_{-\bm k} 
\end{pmatrix}
= 
\begin{pmatrix}
\cosh\psi_{\bm k}^{\alpha} & \sinh\psi_{\bm k}^{\alpha} \\
\sinh\psi_{\bm k}^{\alpha} & \cosh\psi_{\bm k}^{\alpha} 
\end{pmatrix}
\begin{pmatrix}\alpha_{\bm k} \\ \alpha^{\dag}_{-\bm k} 
\end{pmatrix}, \label{eq:bogoliubov_cant2_1} \\
\begin{pmatrix}d_{\bm k} \\ d^{\dag}_{-\bm k} 
\end{pmatrix}
= 
\begin{pmatrix}
\cosh\psi_{\bm k}^{\beta} & \sinh\psi_{\bm k}^{\beta} \\
\sinh\psi_{\bm k}^{\beta} & \cosh\psi_{\bm k}^{\beta}
\end{pmatrix}
\begin{pmatrix}\beta_{\bm k} \\ \beta^{\dag}_{-\bm k} 
\end{pmatrix}, \label{eq:bogoliubov_cant2_2}
\end{align}
where $\psi^{\nu}_{{\bm k}}$ is chosen to satisfy
\begin{align}
\tanh(2\psi^\alpha_{\bm k})&=\frac{C_{\bm k}}{A_{\bm k}}, \\ \tanh(2\psi^\beta_{\bm k})&=\frac{-C_{\bm k}}{B_{\bm k}}.
\end{align}
By dropping a constant term, we finally obtain the following Hamiltonian:
\begin{align}
\mathcal{H}_{\rm CAF} = 
\sum_{\bm k} \varepsilon_{{\bm k}\alpha}\alpha_{\bm k}^\dag \alpha_{\bm k}+
\sum_{\bm k} \varepsilon_{{\bm k}\beta}\beta_{\bm k}^\dag \beta_{\bm k},\label{eq:cant_diag}
\end{align}
where $\varepsilon_{\nu}(\bm k)$ is a dispersion of the two bosonic modes given as
\begin{align}
\varepsilon_{{\bm k}\alpha}&=\Big[ 2J_3S(3-\gamma_{\bm k})\{2J_3S(3+\gamma_{\bm k})\nonumber \\
&\hspace{7mm}+h\cos \phi -4J_3S(3+\gamma_{\bm k})\cos^2\phi \} \Big]^{1/2},\\
\varepsilon_{{\bm k}\beta}&=\Big[2J_3S(3+\gamma_{\bm k})\{2J_3S(1-2\cos^2\phi) (3-\gamma_{\bm k})\nonumber \\
&\hspace{7mm}+h\cos \phi \} \Big]^{1/2}.
\label{Hcantdiag}
\end{align}

\subsection{Interfacial exchange coupling}

We consider the exchange coupling at the interface, with Hamiltonian is given as
\begin{align}
\mathcal{H}_{\rm int} = \sum_{{\bm k},{\bm k}'} \left(
T_{{\bm k},{\bm k}'} {\cal S}^-_{\bm k} S^+_{{\bm k}'}+
T_{{\bm k},{\bm k}'}^* {\cal S}^+_{\bm k}S^-_{{\bm k}'}\right),
\end{align}
where ${\cal S}^\pm_{\bm k}={\cal S}^x_{\bm k} \pm i {\cal S}^y_{\bm k}$ and $S^\pm_{\bm k}=S^x_{\bm k} \pm i S^y_{\bm k}$ are Fourier transformations of the spin ladder operators in the FI and the SN or the canted antiferromagnetic insulator, respectively.
Since we focus on the uniform spin precession in the FI, the coupling at the interface is approximated as
\begin{align}
{\cal H}_{\rm int} = \sqrt{2S_0} \sum_{{\bm k}'} (T_{{\bm 0},{\bm k}'} b^\dagger_{\bm 0} S^+_{{\bm k}'} + T_{{\bm 0},{\bm k}'}^* b_{\bm 0} S^-_{{\bm k}'} ).
\end{align}
For simplicity, we set $T_{{\bm 0},{\bm k}'}=\bar{T}$, assuming a dirty interface~\cite{Kato2019,Ominato2020b,Heydari2025,Yama2025}.
We should note that the spin ladder operators are defined in the laboratory coordinates $(x,y,z)$, which are related to the coordinates $(x',y',z')$ as
\begin{align}
\begin{pmatrix}
S^x_i \\
S^y_i \\
S^z_i 
\end{pmatrix}
= 
\begin{pmatrix}
1 & 0 & 0 \\
0 & \cos \theta & -\sin \theta \\
0 & \sin \theta & \cos \theta
\end{pmatrix}
\begin{pmatrix}
S^{x'}_i \\
S^{y'}_i \\
S^{z'}_i 
\end{pmatrix}.
\label{Stransformation}
\end{align}

\section{Formulation} 
\label{sec:formulation}

In this section, we formulate the Gilbert damping in the FMR experiment for the bilayer system.
First, we briefly describe a general formulation in Sec.~\ref{sec:GeneralFromulation}.
Next, we calculate the increase of the Gilbert damping into the SN and CAF insulators in Sec.~\ref{sec:FormulationSN} and Sec.~\ref{sec:FormulationCant}, respectively.

\subsection{Gilbert damping}
\label{sec:GeneralFromulation}

We first define the temperature Green's function for magnons in the FI as follows:
\begin{align}
G(\tau) &= -\frac{1}{\hbar} \langle S^+_{{\bm 0}}(\tau) S^-_{{\bm 0}}(0) \rangle \nonumber \\
&= -\frac{2S_0}{\hbar} \langle b_{{\bm 0}}(\tau) b^\dagger_{{\bm 0}}(0) \rangle ,\\
G(i\omega_n) &= \int_0^{\hbar\beta} d\tau \, e^{i\omega_n \tau} G(\tau),
\end{align}
where $A(\tau) = e^{{\cal H} \tau} A e^{-{\cal H} \tau}$, $\omega_n = 2\pi nk_{\rm B}T/\hbar$ is the bosonic Matsubara frequency, and $\beta$ is the inverse temperature.
Using linear response theory, we can show that the microwave absorption rate is proportional to ${\rm Im}\, G^R(\omega)$, where $G^R(\omega)=G(i\omega_n\rightarrow \omega+i\delta)$ is a retarded component of the spin correlation function~\cite{Funato2022}.
For an isolated FI, the retarded Green's function is calculated as
\begin{align}
G^R_{0}(\omega)=\frac{2S_0/\hbar}{\omega - \omega_{\bm 0}+i \alpha_{\rm G} },
\end{align}
where we introduce a phenomenological damping constant, $\alpha_{\rm G}$, to describe the Gilbert damping in the bulk FI.

In the presence of exchange coupling between the FI and the spin-nematic insulator, the retarded Green's function is modified as
\begin{align}
G^R(\omega)=\frac{2S_0/\hbar}{\omega - \omega_{\bm 0} +i \alpha_{\rm G} - \Sigma^R(\omega)},
\end{align}
where $\Sigma^R(\omega)$ is the retarded component of the self-energy due to the interface, which is obtained by the analytic continuation $\Sigma^R(\omega) = \Sigma(i\omega_n \rightarrow \omega + i\delta)$ from the self-energy in the imaginary-time formalism, $\Sigma(i\omega_n)$.
Assuming the FMR peak is sufficiently sharp, the increase of the Gilbert damping is well approximated as
\begin{align}
\delta \alpha_{\rm G}
= -\frac{2S_0}{\hbar\omega_{\bm 0}} {\rm Im} \, \Sigma^R(\omega_{\bm 0}).
\label{GDexpression}
\end{align}
We note that the real part of the self-energy corresponds to the shift of the FMR frequency~\cite{Yama2023a}.
Although this change in FMR frequency is expected to include information on the spin-nematic phase, we only focus on the modulation of the Gilbert damping for simplicity in the discussion.

Within the second-order perturbation with respect to ${\cal H}_{\rm int}$, the self-energy in the imaginary-time formalism is calculated as
\begin{align}
\Sigma(i\omega_n) &= \int_0^{\hbar\beta} d\tau \, e^{i\omega_n \tau} \Sigma(\tau), \\
\Sigma(\tau)&=-\sum_{{\bm k},{\bm k}'} \frac{T_{
{\bm 0},{\bm k}}T_{{\bm 0},{\bm k}'}^*}{\hbar}
\langle S^+_{{\bm k}}(\tau)
S^-_{{\bm k}'}(0) \rangle,
\label{eq_self_energy0}
\end{align}
where the average is taken for thermal equilibrium of the spin-nematic state. 
Using the coordinate transformation given in Eq.~(\ref{Stransformation}), the self-energy is obtained as
\begin{align}
\Sigma(\tau)&=-\sum_{a,b=1}^3 \sum_{{\bm k},{\bm k}'} \frac{T_{{\bm 0},{\bm k}}T_{{\bm 0},{\bm k}'}^*}{\hbar}
g_a(\theta) g_b^*(\theta) \nonumber \\
&\hspace{6mm} \times \left\langle S^{(a)}_{{\bm k}}(\tau)
\left\{S^{(b)}_{{\bm k}'}(0)\right\}^\dag \right\rangle,
\label{eq_self_energy}
\end{align}
where
\begin{align}
& {S}^{(1)}_{\bm{k}}={S}^{z'}_{\bm{k}},
\quad g_{1}(\theta)=-i\sin{\theta}, \\
& {S}^{(2)}_{\bm{k}}={S}^{+'}_{\bm{k}},
\quad g_{2}(\theta)=\cos^2(\theta/2), \\
& {S}^{(3)}_{\bm{k}}={S}^{-'}_{\bm{k}},
\quad g_{3}(\theta)=\sin^2 (\theta/2). 
\end{align}

\subsection{Spin pumping into SN insulator}
\label{sec:FormulationSN}

For the model of the SN state, the self-energy is written as
\begin{align}
\Sigma(\tau)&=-\sum_{a,b=1}^3 \sum_{{\bm k},{\bm k}'} \frac{|\bar{T}|^2}{\hbar}
g_a(\theta) g_b^*(\theta) \left\langle S^{(a)}_{{\bm k}}(\tau)
\left\{S^{(b)}_{{\bm k}'}(0)\right\}^\dag \right\rangle .
\label{eq_self_energy_dirty}
\end{align}
Using the imaginary-time spin-spin correlation function
\begin{align}
\chi^{\mu\nu}_{{\bm k},{\bm k}'}(\tau) = -\frac{1}{\hbar} \left\langle S^\mu_{{\bm k}}(\tau)
S^\nu_{{\bm k}'}(0) \right\rangle, \quad (\mu,\nu=z',+',-'),
\end{align}
the self-energy is rewritten as
\begin{align}
\Sigma(\tau)&= \sum_{{\bm k}} \frac{|\bar{T}|^2}{\hbar} \Bigl\{ \sin^2 \theta \, \chi^{z'z'}_{{\bm k},-{\bm k}}(\tau) \nonumber \\
&+ \cos^4\left(\frac{\theta}{2}\right) \chi^{+'-'}_{{\bm k},-{\bm k}}(\tau) + \sin^4\left(\frac{\theta}{2}\right) \chi^{-'+'}_{{\bm k},-{\bm k}}(\tau) \nonumber \\
& + \frac{\sin^2 \theta}{4} [ \chi^{+'+'}_{{\bm k},-{\bm k}}(\tau)
+ \chi^{-'-'}_{{\bm k},-{\bm k}}(\tau) ] \Bigr\},
\end{align}
where we have used translational invariance of ${\cal H}_{\rm SN}$ and $\chi^{z'+'}_{{\bm k},{\bm k}'} = \chi^{z'-'}_{{\bm k},{\bm k}'}= \chi^{+'z'}_{{\bm k},{\bm k}'}= \chi^{-'z'}_{{\bm k},{\bm k}'}=0$.
Here, it is remarkable that 
$\chi^{-'-'}_{{\bm k},-{\bm k}}(\tau)$ and $\chi^{+'+'}_{{\bm k},-{\bm k}}(\tau)$ do not vanish as a consequence of the squeezing terms in Eq. (14).

The remaining task is to calculate the spin-spin correlation function $\chi_{{\bm k},-{\bm k}}^{\mu\nu}(\tau)$ and its Fourier transformation.
This is done by using the mean-field Hamiltonian, Eq.~(\ref{HSNdiag}), written in terms of the Schwinger bosons.
The detailed calculation is given in Appendix~\ref{app:spin_sus_nematic}. 
Consequently, the increase of the Gilbert damping for $\omega_{\bm 0}>0$ is given as
\begin{align}
\delta \alpha_{\rm G}&= \frac{2\pi S_0|\bar{T}|^2}{\hbar\omega_{\bm 0}} \sum_{\bm k}\sum_{\nu = z,\perp} C_\nu \delta(\hbar\omega_{\bm 0}-\varepsilon_{{\bm k}\nu}),
\label{nematic_result1}
\end{align}
where
\begin{align}
C_z &=M \Bigl[ \sin \mu \cos \theta \nonumber \\
&+  \{ \cosh 2\xi_{{\bm k}z}  - \cos \mu \sinh 2\xi_{{\bm k}z}\} \cos^2\theta
\nonumber \\
&+ \frac{1+\cos\mu}{2} \{\cosh 2\xi_{{\bm k}z}-\sinh 2\xi_{{\bm k}z}\} \sin^2 \theta \Bigr], \label{nematic_result2} \\
C_\perp &=M \, \cos^2 \mu \, [ \cosh (2\xi_{{\bm k}\perp}) - \sinh (2\xi_{{\bm k}\perp}) ] \sin^2 \theta.
\label{nematic_result3}
\end{align}

\subsection{Spin pumping into CAF insulator}
\label{sec:FormulationCant}

Similarly, the self-energy is calculated for the model of the CAF insulator as
\begin{align}
\Sigma(\tau)&= \sum_{{\bm k}} \sum_{{\bm k'}=\pm{\bm k}} \frac{|\bar{T}|^2}{\hbar} \Bigl\{ \sin^2 \theta \, \chi^{z'z'}_{{\bm k},{\bm k'}}(\tau) \nonumber \\
&+ \cos^4\left(\frac{\theta}{2}\right) \chi^{+'-'}_{{\bm k},{\bm k'}}(\tau) + \sin^4\left(\frac{\theta}{2}\right) \chi^{-'+'}_{{\bm k},{\bm k'}}(\tau) \nonumber \\
& + \frac{\sin^2 \theta}{4} [ \chi^{+'+'}_{{\bm k},{\bm k'}}(\tau)
+ \chi^{-'-'}_{{\bm k}, {\bm k'}}(\tau) ] \Bigr\},
\label{eq:selfenergy_cant}
\end{align}
where $\chi^{\mu\nu}_{{\bm k},{\bm k'}}(\tau)$ ($\mu,\nu=z',+',-'$) is an imaginary-time spin-spin correlation function
\begin{align}
\chi^{\mu\nu}_{{\bm k},{\bm k'}}(\tau) = - \frac{1}{\hbar} \sum_{\eta = {\rm A}, {\rm B}}  \left\langle S^\mu_{{\eta, \bm k}}(\tau)
S^\nu_{{\eta, \bm k}'}(0) \right\rangle. 
\end{align}
We have used translational invariance of ${\cal H}_{\rm CAF}$ and $\chi^{z'+'}_{{\bm k},{\bm k}'} = \chi^{z'-'}_{{\bm k},{\bm k}'}= \chi^{+'z'}_{{\bm k},{\bm k}'}= \chi^{-'z'}_{{\bm k},{\bm k}'}=0$.
We note that $\chi^{-'-'}_{{\bm k},-{\bm k}}(\tau)$ and $\chi^{+'+'}_{{\bm k},-{\bm k}}(\tau)$ do not vanish as in the SN insulator, reflecting the absence of the U(1) symmetry.
The spin-spin correlation function $\chi_{{\bm k},{\bm k'}}^{\mu\nu}(\tau)$ is obtained from the diagonalized Hamiltonian, Eq.~(\ref{Hcantdiag}).
Consequently, the increase of the Gilbert damping for $\omega_{\bm 0}>0$ is given as
\begin{align}
\delta \alpha_{\rm G}&= \frac{2\pi S_0|\bar{T}|^2}{\hbar\omega_{\bm 0}} \sum_{\bm k}\sum_{\nu = \alpha,\beta} C_\nu \delta(\hbar\omega_{\bm 0}-\varepsilon_{{\bm k}\nu}),
\label{canted_result1}\\
C_\nu &=M \Bigl[ \cos\phi \cos \theta \nonumber \\
&  + \frac12 \left\{ (\cos^2\phi+\sin^2\phi \sin^2\theta) (\cosh 2\psi_{\bm k}^\nu + \sinh 2\psi_{\bm k}^\nu) \right\} \nonumber \\
& +\frac12 \left\{(1 -\sin^2 \theta) (\cosh 2\psi_{\bm k}^\nu - \sinh 2\psi_{\bm k}^\nu) \right\} \Bigr] .\label{canted_result3}
\end{align}
A detailed calculation is given in Appendix~\ref{app:spin_sus_cant}. 

\section{Results}
\label{sec:Result}

In this section, we present the results for the increase of the Gilbert damping in the FI/SN and FI/CAF bilayer systems.
First, we show the results in the absence of a magnetic field ($h = 0$) and under a finite magnetic field ($h > 0$) in Sec.~\ref{sec:ResultNoMagneticField} and Sec.~\ref{sec:Result_magnetic_field}, respectively.
Throughout this paper, the parameters are set as $J_1=0$, $J_2=-{\cal J}$, and $J_3={\cal J}$, where this choice of $J_1$ and $J_2$ realizes the spin nematic phase.
Furthermore, we replace the delta-function in Eqs.~(\ref{nematic_result1}) and (\ref{canted_result1}) with the Lorentzian function of width $0.03{\cal J}$ to plot the dependence on the FMR frequency $\omega_{\bm 0}$.
We note that $\delta \alpha_{\rm G}$ becomes temperature-independent within the mean-field theory, which is justified at low temperatures.

\subsection{Zero magnetic field}
\label{sec:ResultNoMagneticField}

\begin{figure}[tb]
\begin{center}
\includegraphics[clip,width=8.0cm]{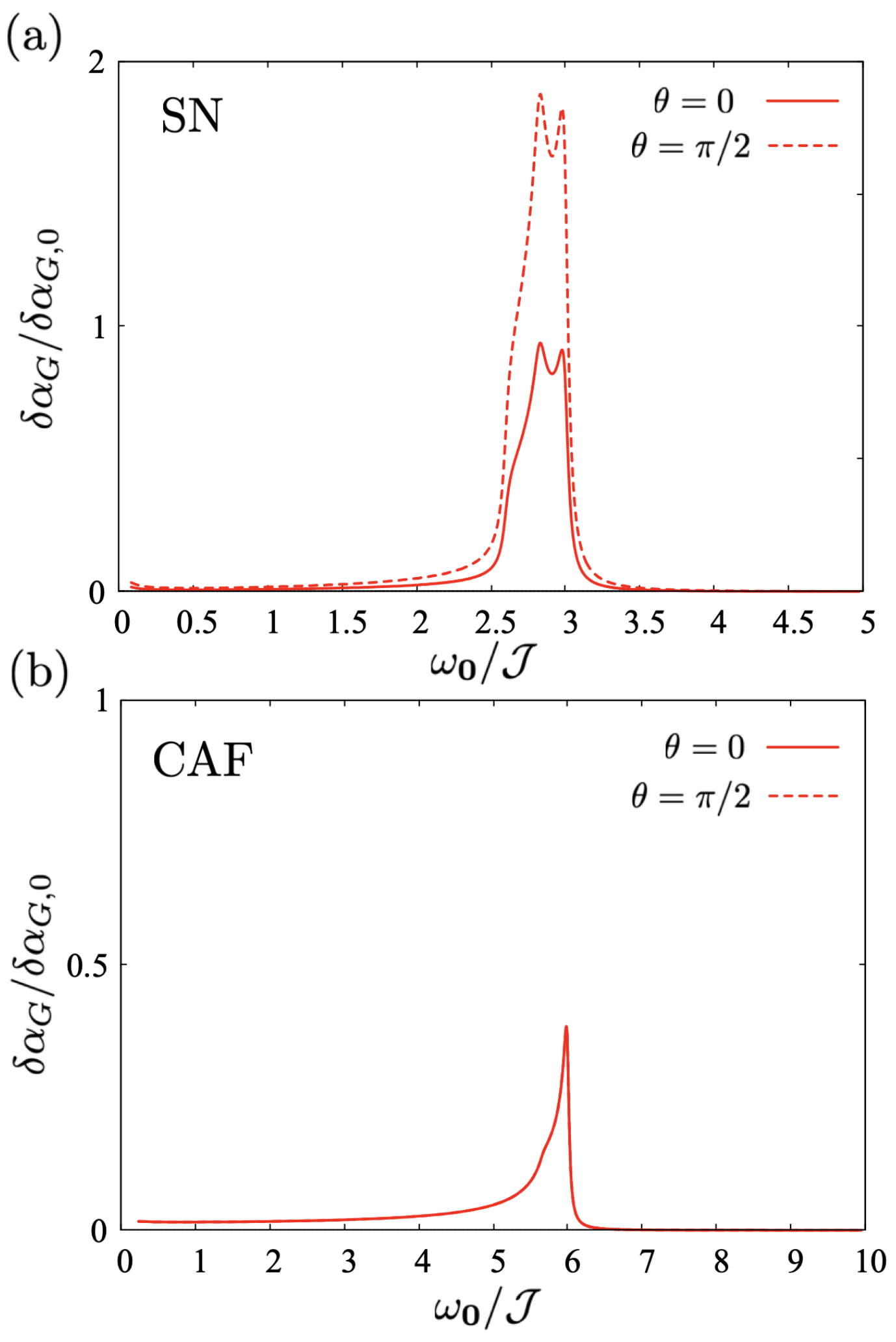}
\caption{
The increase of the Gilbert damping as a function of the FMR frequency $\omega_{\bm 0}$ for (a) the SN insulator and (b) the CAF insulator.
The solid and dashed lines represent $\theta=0$ and $\theta=\pi/2$, respectively. The parameters are set as $J_1=0$, $J_2=-{\cal J}$, $J_3={\cal J}$ and $h=0$.
We note that the Gilbert damping is independent of $\theta$ for the CAF insulator.}
\label{fig:h0}
\end{center}
\end{figure}

In Fig.~\ref{fig:h0}~(a) and (b), we plot $\delta \alpha_{\rm G}/\delta \alpha_{{\rm G},0}$ as a function of the FMR frequency $\omega_{\bm 0}$ for the SN insulator and the CAF insulator, respectively.
Here, $\delta \alpha_{{\rm G},0} = 2\pi S_0|\bar{T}|^2 /{\cal J}^2$ is a normalization factor.
We note that in the absence of the magnetic field ($h=0$), the model for the CAF insulator leads to collinear AF ordering (no spin canting).
For the SN state, the enhancement of Gilbert damping for $\theta = \pi/2$ is twice that of $\theta = 0$.
In general, $\delta \alpha_{\rm G}$ is proportional to $1+\sin^2 \theta$. 
This dependence arises from the fact that $\delta \alpha_{\rm G}$ is a sum of the two contributions from two types of Schwinger bosons; the contribution of the $z'$-component Schwinger bosons is independent of $\theta$, while that of the $x'$-component Schwinger bosons is proportional to $\sin^2\theta$ (see also Eqs.~(\ref{nematic_result1})-(\ref{nematic_result3})).
In contrast, there is no angle dependence for the antiferromagnetic insulator as shown in Fig.~\ref{fig:h0}~(b). This corresponds to the fact that, in the absence of a magnetic field, only a collinear antiferromagnetic order is established, exhibiting rotational symmetry around the $x'$-axis.

\subsection{Finite magnetic field}
\label{sec:Result_magnetic_field}

\begin{figure}[tb]
\begin{center}
\includegraphics[clip,width=8.0cm]{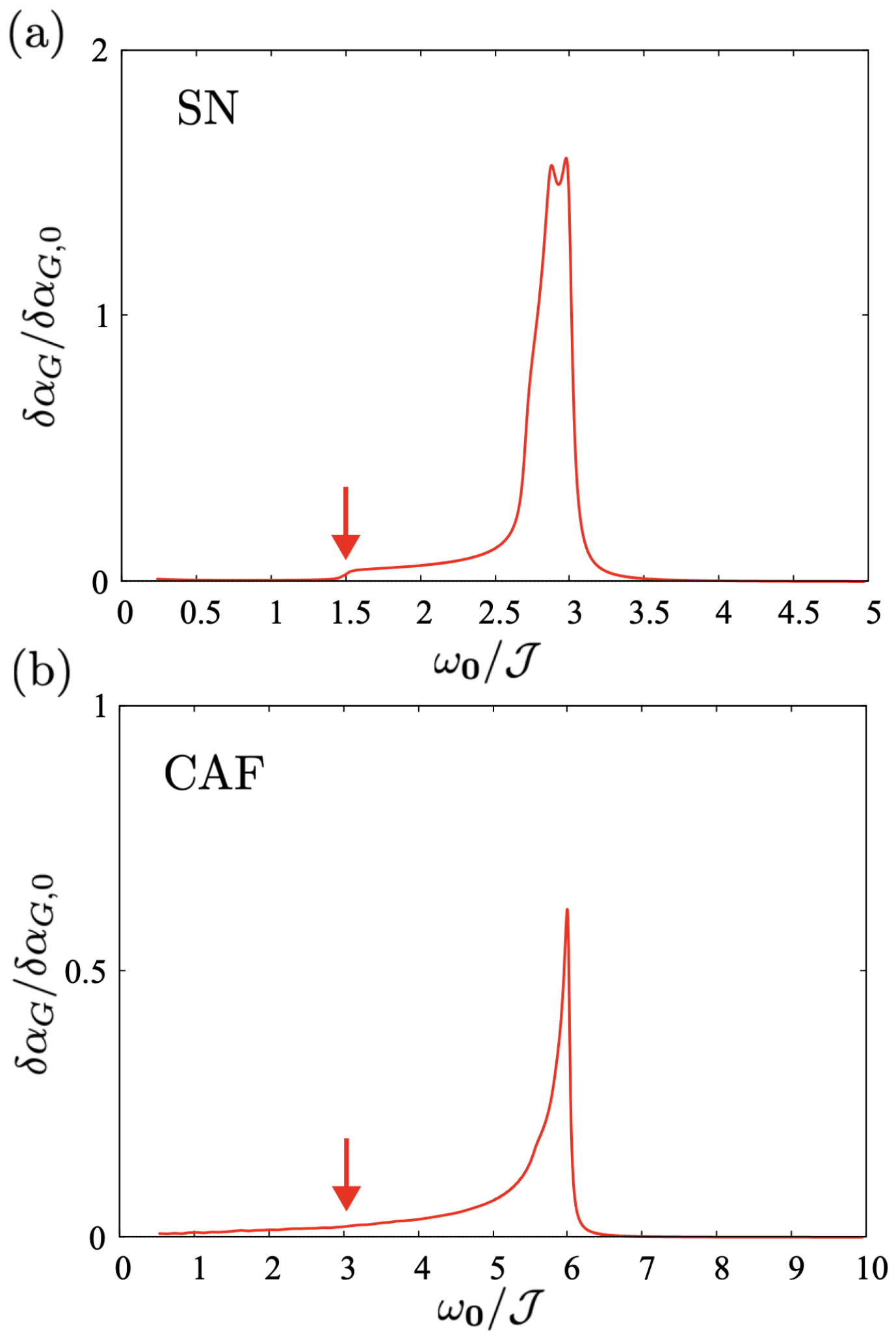}
\caption{
The increase of the Gilbert damping as a function of the FMR frequency $\omega_{\bm 0}$ at $\theta=0$ for (a) the SN insulator and (b) the CAF insulator. 
The parameters are set as $J_1=0$, $J_2=-{\cal J}$, and $h=3{\cal J}$. Red arrows indicate the energy gap.}
\label{fig:h3}
\end{center}
\end{figure}

\begin{figure}[tb]
\begin{center}
\includegraphics[clip,width=8.0cm]{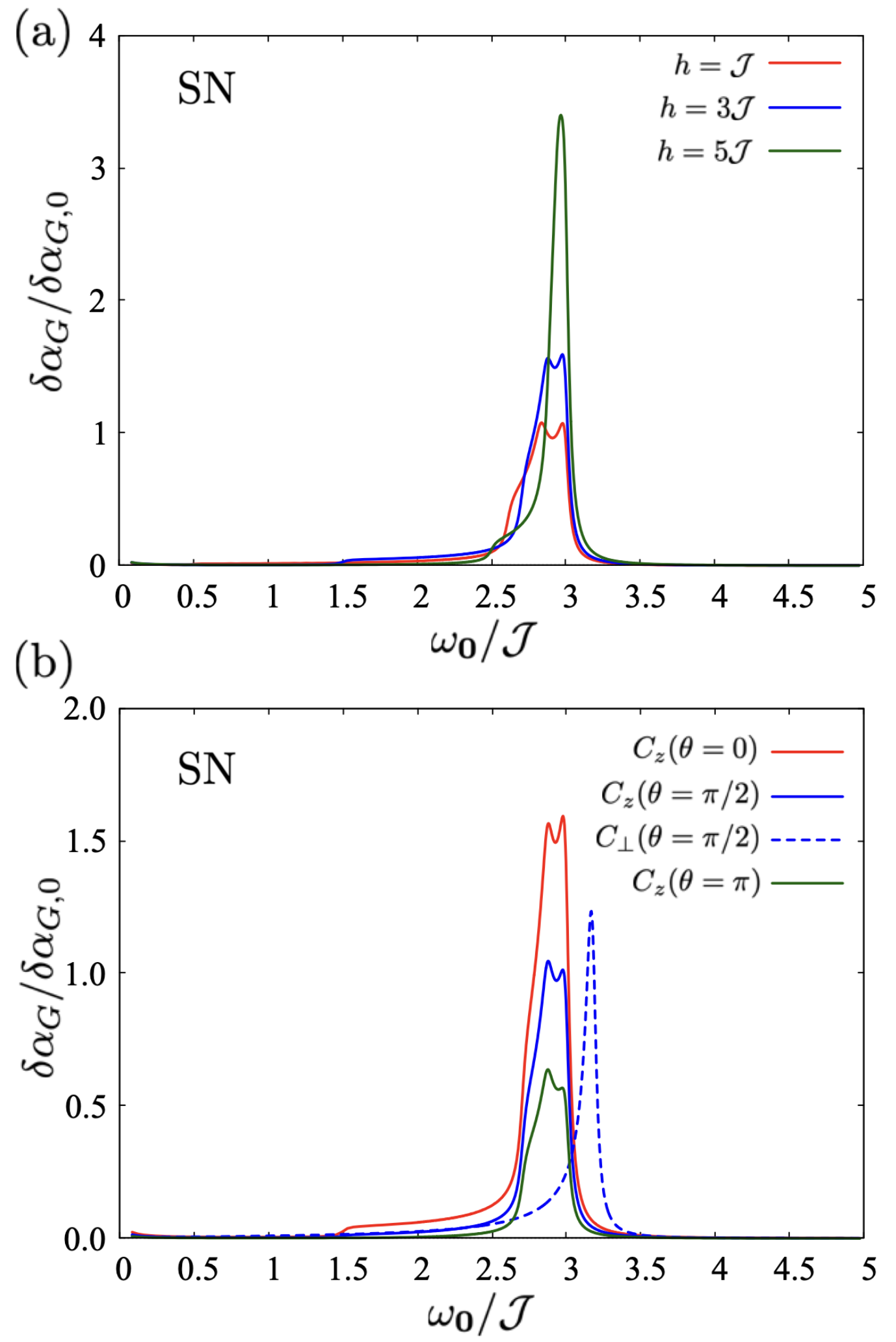}
\caption{The increase of the Gilbert damping for the SN case is shown as a function of the FMR frequency $\omega_{\bm 0}$ for $J_1=0$ and $J_2=-{\cal J}$. (a) The magnetic-field dependence at $\theta=0$. The three curves correspond to $h={\cal J}$, $3{\cal J}$, and $5{\cal J}$, respectively. (b) The angle dependence at $h=3{\cal J}$. The red, blue, and green solid lines correspond to the contribution of $C_z$ for $\theta=0$, $\theta=\pi/2$, and $\theta=\pi$, respectively. The dashed line represents the contribution of $C_\perp$ at $\theta = \pi/2$. We note that $C_\perp (\theta=0) = C_\perp (\theta=\pi) = 0$.}
\label{fig:nem_angle}
\end{center}
\end{figure}

Fig.~\ref{fig:h3} shows $\delta \alpha_{\rm G}/\delta \alpha_{{\rm G},0}$ as a function of the FMR frequency $\omega_{\bm 0}$ in the presence of an external magnetic field at $\theta=0$. For the SN insulator, $\delta \alpha_{\rm G}$ becomes zero at frequencies below the energy gap (indicated by the red arrow), as shown in Fig.~\ref{fig:h3}~(a). However, for the CAF insulator, a finite value remains, as shown in Fig.~\ref{fig:h3}~(b). This is because, at $\theta=0$, only the gapped bosons contribute to the spin pumping in the SN case, whereas both gapped and gapless bosons contribute in the CAF case (see also Eqs.~(\ref{nematic_result1})-(\ref{nematic_result3}) and Eqs.~(\ref{canted_result1})-(\ref{canted_result3})).

\begin{figure*}[tb]
\begin{center}
\includegraphics[clip,width=17.0cm]{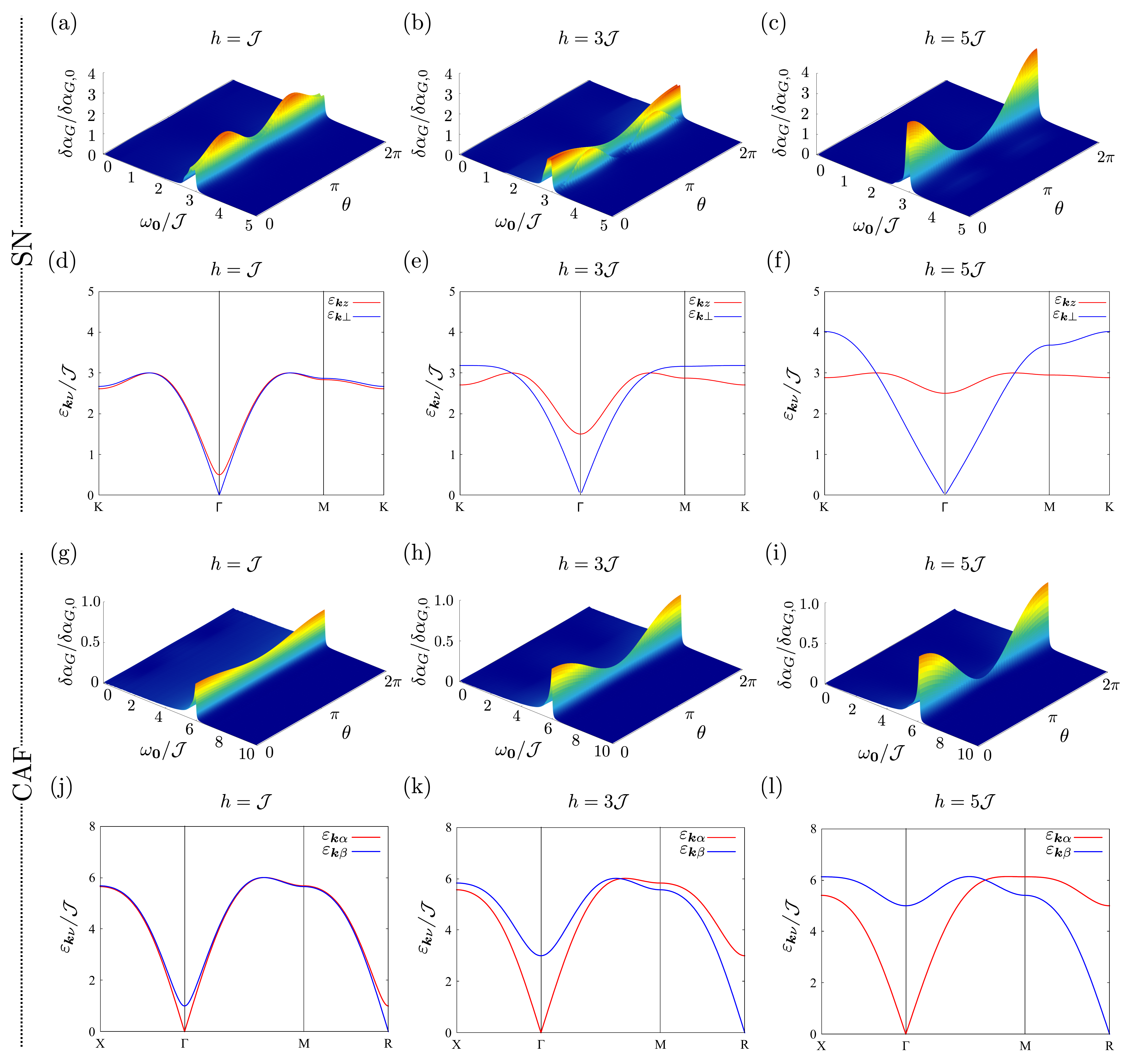}
\caption{
(a)-(c) The color gradient in the three-dimensional plot represents the increase of the Gilbert damping, $\delta \alpha_{\rm G}$, of the SN/FI junction as a function of the FMR frequency $\omega_{\bm 0}$ and $\theta$ for $h={\cal J}$,  $3{\cal J}$, and $5{\cal J}$, respectively.
(d)-(f) The corresponding dispersion of the two Schwinger bosons of $\varepsilon_{{\bm k} z}$ and $\varepsilon_{{\bm k} \perp}$.
(g)-(i) The color gradient in the three-dimensional plot represents the increase of the Gilbert damping, $\delta \alpha_{\rm G}$, of the CAF/FI junction for $h={\cal J}$, $3{\cal J}$, and $5{\cal J}$, respectively.
(j)-(l) The corresponding dispersion of the two magnon modes,  $\varepsilon_{{\bm k} \alpha}$ and $\varepsilon_{{\bm k} \beta}$.
}
\label{fig:x_anisotropy}
\end{center}
\end{figure*}

To clarify the magnetic field dependence in the SN case, we 
show $\delta \alpha_{\rm G}/\delta \alpha_{{\rm G},0}$ as a function of the FMR frequency $\omega_{\bm 0}$ in the presence of an external magnetic field in Fig.~\ref{fig:nem_angle}.
As the magnetic field increases, $\delta \alpha_{{\rm G}}(\omega_{\bm 0})$ grows because the net magnetization in the spin-nematic insulator increases and the peak becomes sharper reflecting changes in the dispersion relation of the Schwinger bosons
(see also Fig.~\ref{fig:x_anisotropy}(d)-(f)).
Fig.~\ref{fig:nem_angle}~(b) shows the increase of the Gilbert damping at $\theta=0$, $\theta=\pi/2$, $\theta=\pi$ with the magnetic field fixed at $h=3{\cal J}$, where the contributions of the two kinds of Schwinger bosons are plotted separately.
We note that the contribution from the perpendicular Schwinger bosons vanishes for $\theta = 0, \pi$.
The peak at $\omega_{\bm 0}=2.9{\cal J}$ is the contribution of the $z'$-component Schwinger bosons, and the peak at $\omega_{\bm 0}=3.2{\cal J}$ is that of the perpendicular component of the Schwinger bosons.
As seen in the figure, the two components of the bosons have different dependencies on $\theta$.

To compare the dependence on the magnetic-field orientation between the SN/FI and CAF/FI junctions, we show the increase of the Gilbert damping as a function of $\omega_{\bm 0}$ and $\theta$ for $h = {\cal J}$, $3{\cal J}$, and $5{\cal J}$ in Fig.~\ref{fig:x_anisotropy}(a)-(c) and (g)-(i), where the first column (third column) corresponds to the SN/FI (CAF/FI) junction.
For both junctions, the peak height at a finite frequency varies as the angle $\theta$ changes.
The shape and position of the peak of $\delta \alpha_{\rm G}$ are determined by the corresponding dispersion of the two bosons, as shown in Fig.~\ref{fig:x_anisotropy}(d)-(f) for the SN case and (j)-(l) for the CAF case.
As expected from the form of Eq.~(\ref{nematic_result1}) and Eq.~(\ref{canted_result1}), the peak frequency of $\delta \alpha_{\rm G}$ corresponds to the energy of the flat part in the dispersion of the two bosons in both cases.
The low-frequency shoulder structure of $\delta \alpha_{\rm G}$ reflects the low-energy states of the Schwinger bosons, though its weight is much weaker than that of the main peak.

Let us examine the detailed features of the SN/FI junction under a finite magnetic field. 
In Fig.~\ref{fig:x_anisotropy}(a)-(c), we show three-dimensional plots of $\delta \alpha_{\rm G}$ as a function of the resonant frequency $\omega_{\bm 0}$ and the angle $\theta$ for the SN/FI junction.
For comparison, we also show the same plots for the CAF/FI junction in Fig.~\ref{fig:x_anisotropy}(g)-(i).
For $h={\cal J}$ (Fig.~\ref{fig:x_anisotropy}(a)), $\delta \alpha_{\rm G}$ has a single peak around $\omega_{\bm 0}\simeq 3{\cal J}$ and
this peak becomes most significant near $\theta = \pi/2, 3\pi/2$.
For a higher magnetic field of $h=3{\cal J}$ (Fig.~\ref{fig:x_anisotropy}(b)), a secondary peak appears in the high-energy side with a different $\theta$-dependence; the lower-energy peak around $\omega_{\bm 0}=2.9{\cal J}$ becomes significant near $\theta = 0$, while the high-energy peak around $\omega_{\bm 0}=3.2{\cal J}$ appears around $\theta = \pi/2,3\pi/2$.
The positions of the double peaks correspond to the flat bands, whose energies are different for the two types of bosons (see Fig.~\ref{fig:x_anisotropy}(e)), while their weights have different dependencies on $\theta$, as seen also in Eqs.~(\ref{nematic_result1})-(\ref{nematic_result3}).
For the highest magnetic field of $h=5{\cal J}$ (Fig.~\ref{fig:x_anisotropy}(c)), the contribution of the perpendicular component of the Schwinger bosons almost vanishes, while that of the $z'$-component bosons becomes dominant. As a result, $\delta \alpha_{\rm G}$ has only a single peak due to the flat part in the dispersion of the $z'$ Schwinger bosons.
For the CAF/FI junction (Fig.~\ref{fig:x_anisotropy}(g)-(i)), there always appears a single peak regardless of the magnetic field strength, whose height has a maximum value  at $\theta = 0, 2\pi$.
It is remarkable that no peaks are observed near $\theta = \pi/2,3\pi/2$ as in the SN/FI junction.

\section{Summary}
\label{sec:Summary}

In this study, we examined spin pumping into a spin-nematic (SN) state in a junction system composed of a ferromagnetic insulator and a spin-nematic insulator.
To clarify its qualitative features, we employed a spin-1 bilinear-biquadratic model to realize the spin-nematic phase and calculated the increase of the Gilbert damping using the mean-field theory based on the Schwinger boson formalism. 
To clarify the feature of the SN state, we compared the obtained results for the model of the canted antiferromagnetic insulator (CAF) because both the SN and CAF states have two bosonic modes and do not conserve a spin component in a specific direction due to the U(1) symmetry breaking.
Although these two models lead to similar results at first glance, there are clear differences between them.
(i) For zero magnetic field, $\delta \alpha_{\rm G}$ depends on the spin angle of the FI, $\theta$, for the SN/FI junction, while it is independent of $\theta$ for the CAF/FI junction (see Fig.~\ref{fig:h0}).
(ii) For a finite magnetic field, $\delta \alpha_{\rm G}$ vanishes below a specific frequency for the SN/FI junction, while it continues to be finite for the CAF/FI junction (see Fig.~\ref{fig:h3}).
(iii) For a finite magnetic field, the peaks in $\delta \alpha_{\rm G}$ have a different $\theta$-dependence between the SN/FI and CAF/FI junctions (see Fig.~\ref{fig:x_anisotropy}); 
For the CAF/FI junction, there appears a single peak with $\cos \theta$ type dependence, while for the SN/FI junction another peak with  $\cos 2\theta$ type dependence coexists when the magnetic field is not too strong.
These findings are also consistent with analytic results in Sec.~\ref{sec:formulation}.

Our study demonstrates that spin pumping, a well-established technique in spintronics, provides a promising route to identify the spin-nematic (SN) phase. By tuning the microwave frequency, this approach allows systematic control of the relevant energy scale, while altering the orientation of the applied magnetic field enables clear separation of multiple magnetic excitations. Such features are particularly advantageous for probing the unique excitation spectra characteristic of the SN phase. 

Finally, we comment on the experimental realization. The exchange interactions of the candidate materials, ${\mathrm{LiCuVO}}_{4}$ and ${\mathrm{BaCdVO}}(\mathrm{PO}4)_{2}$, for spin nematic states are of the order of $0.5$-$5\,{\rm meV}$, which corresponds to the frequency of $0.1$-$1\, {\rm THz}$.
This frequency scale is much larger than that used in the usual FMR experiment.
Therefore, to observe the main peaks at the frequency scale of ${\cal J}$, we need to use spin systems with smaller exchange interactions or to apply high-speed THz spin pumping technique~\cite{Mizukami2016,Walowski2016}.
On the other hand, the absence of the increase of the Gilbert damping at low frequencies (see Fig.~\ref{fig:h3}(a)) may be observed using the usual FMR experiment.
Study on detailed conditions for observation in realistic candidates of the SN phase is left as future problems.

\section*{Acknowledgements}

The authors thank M. Matsuo and M. Sato for their helpful discussions. T. I. is supported by the International Graduate Program of Innovation for Intelligent World (IIW) of the University of Tokyo and the International Program of the Institute for Solid State Physics, the University of Tokyo and acknowledges the support and hospitality of the IRTG Fluctuations and Nonlinearities of SFB 1432 during his stay at University of Konstanz. T. K. acknowledges the support of the Japan Society for the Promotion of Science (JSPS KAKENHI Grant No.~JP24K06951). W. B. acknowledges financial support from the Deutsche Forschungsgemeinschaft (DFG, German Research Foundation) through Project-ID 425217212 - SFB 1432.

\appendix

\section{Detailed calculation of the spin-spin correlation function}
\label{app:spin_sus_nematic}

In this appendix, we provide a detailed calculation of the spin-spin correlation functions.
We first rewrite the spin operators in the $(x',y',z')$ coordinate system using the Schwinger bosons, following Eqs.~(\ref{SxdashBoson})-(\ref{SzdashBoson}). We then apply the transformation given in Eq.~(\ref{bosontrans}) and the condensation condition for the parallel component boson, given in Eq.~(\ref{condensepara}).
As a result, the spin operators are calculated as
\begin{widetext}
\begin{align}
S^{+'}_i
&=i\sqrt{M}\left[ \left( \cos \frac{\mu}{2}- \sin \frac{\mu}{2}\right) a_{z'}^\dag(i)- \left(\cos \frac{\mu}{2}+ \sin \frac{\mu}{2} \right) a_{z'}(i) \right] , \\
S^{z'}_i&=M\sin \mu +i\sqrt{M} \cos \mu  \, [a_{\perp}(i)- a^\dag_{\perp}(i) ],
\end{align}
and $S_i^{-'} = (S_i^{+'})^\dagger$.
Here, we have omitted higher-order terms with respect to $1/M$.
By performing the Fourier transformation and the Bogoliubov transformation given in Eqs.~(\ref{Bogoliubov1}) and (\ref{Bogoliubov2}), we obtain
\begin{align}
S^{+'}_{\bm k}&=i\sqrt{M} \, \bigl [ (C^-_{\mu} c^z_{{\bm k}}-C^+_{\mu}s^z_{{\bm k}}) \alpha_{{\bm k}z}^\dag +(C^-_{\mu} s^z_{{\bm k}}-C^+_{\mu}c^z_{{\bm k}}) \alpha_{-{\bm k}z}  \bigr] ,\\
S^{z'}_{\bm k}&=M\sin \mu +i\cos \mu \sqrt{M} \, (c_{\bm k}^{\perp}-s_{\bm k}^{\perp})[ (\alpha_{-{\bm k}\perp} -\alpha^\dag_{{\bm k} \perp})] ,
\end{align}
and $S_{\bm k}^{-'} = (S_{-{\bm k}}^{+'})^\dagger$, where $C_\mu^{\pm} = \cos(\mu/2) \pm \sin(\mu/2)$, $c^\nu_{{\bm k}} = \cosh \xi_{{\bm k}\nu}$, and $s^\nu_{{\bm k}} = \sinh \xi_{{\bm k}\nu}$.
Using these operators, the spin-spin correlation functions are calculated as
\begin{align}
\chi^{+'-'}_{{\bm k},-{\bm k}}(\tau)&=M\bigl[(C^-_{\mu} s^z_{{\bm k}}-C^+_{\mu}c^z_{{\bm k}})^2 G_{{\bm k}z}(\tau) + (C^-_{\mu} c^z_{{\bm k}}-C^+_{\mu}s^z_{{\bm k}})^2  G_{{\bm k}z}(-\tau)\bigr], \\
\chi^{-'+'}_{{\bm k},-{\bm k}}(\tau)&=M\bigl[(C^-_{\mu} c^z_{{\bm k}}-C^+_{\mu}s^z_{{\bm k}})^2 G_{{\bm k}z}(\tau) + (C^-_{\mu} s^z_{{\bm k}}-C^+_{\mu}c^z_{{\bm k}})^2 G_{{\bm k}z}(-\tau)\bigr], \\
\chi^{+'+'}_{{\bm k},-{\bm k}}(\tau)&=\chi^{-'-'}_{{\bm k},-{\bm k}}(\tau)=
-M\bigl[ (C^-_{\mu} c^z_{{\bm k}}-C^+_{\mu}s^z_{{\bm k}}) (C^-_{\mu} s^z_{{\bm k}}-C^+_{\mu}c^z_{{\bm k}}) \bigr]\bigl[ G_{{\bm k}z}(\tau)+ G_{{\bm k}z}(-\tau) \bigr], \\
\chi^{z'z'}_{{\bm k},-{\bm k}}(\tau)&=M\cos^2\mu (c_{\bm k}^{\perp}-s_{\bm k}^{\perp})^2 \bigl[ G_{{\bm k}\perp}(\tau)+ G_{{\bm k}\perp}(-\tau) \bigr],
\end{align}
where $G_{{\bm k}\nu}(\tau)$ is a propagator of the Schwinger bosons defined by
\begin{align}
G_{{\bm k}\nu}(\tau) = - \frac{1}{\hbar} \langle \alpha_{{\bm k}\nu}(\tau) \alpha^\dagger_{{\bm k}\nu}(0) \rangle.
\end{align}
Using the diagonalized Hamiltonian, Eq.~(\ref{HSNdiag}), the Fourier transformation of this propagator is calculated as
\begin{align}
G_{{\bm k}\nu}(i\omega_n) = \int_0^{\hbar \beta} d\tau\, e^{i\omega_n \tau} G_{{\bm k}\nu}(\tau)
= \frac{1}{i\hbar \omega_n - \varepsilon_{{\bm k}\nu}} .
\end{align}
Combining these results with Eq.~(\ref{GDexpression}), it is straightforward to derive the final result for $\omega_{\bm 0}>0$, which is
given in Eqs.~(\ref{nematic_result1})-(\ref{nematic_result3}).

\section{Detailed calculation of the spin-spin correlation function in canted antiferromagnetic insulator}
\label{app:spin_sus_cant}

In this appendix, we provide a detailed calculation of the spin-spin correlation functions for the CAF insulator.
We first rewrite the spin operators in the $(x',y',z')$ coordinate system using the magnon operators, following Eqs.~(\ref{eq:sa_cant}) and (\ref{eq:sb_cant}). We then apply the Bogoliubov transformation given in Eqs.~(\ref{eq:bogoliubov_cant1}), (\ref{eq:bogoliubov_cant2_1}) and (\ref{eq:bogoliubov_cant2_2}). As a result, the spin operators are calculated as 
\begin{align}
S^{+'}_{A,\bm k}&=\frac{\sqrt{S}}{2}(\cos\phi+1) (\cosh \psi_{\bm k}^\alpha \alpha_{\bm k}+\sinh \psi_{\bm k}^\alpha \alpha^\dag_{-\bm k}+\cosh \psi_{\bm k}^\beta \beta_{-\bm k}+\sinh \psi_{\bm k}^\beta \beta^\dag_{\bm k}) \nonumber \\
&\hspace{5mm}+\frac{\sqrt{S}}{2}(\cos\phi-1) (\sinh \psi_{\bm k}^\alpha \alpha_{\bm k}+\cosh \psi_{\bm k}^\alpha \alpha^\dag_{-\bm k}+\sinh \psi_{\bm k}^\beta \beta_{-\bm k}+\cosh \psi_{\bm k}^\beta \beta^\dag_{\bm k})-S \sin\phi ,
\\
S^{z'}_{A,\bm k}&=\frac{\sqrt{S}\sin\phi}{2} (\cosh \psi_{\bm k}^\alpha \alpha_{\bm k}+\sinh \psi_{\bm k}^\alpha \alpha^\dag_{-\bm k}+\cosh \psi_{\bm k}^\beta \beta_{-\bm k}+\sinh \psi_{\bm k}^\beta \beta^\dag_{\bm k}) \nonumber \\
&\hspace{5mm}+\frac{\sqrt{S}\sin\phi}{2} (\sinh \psi_{\bm k}^\alpha \alpha_{\bm k}+\cosh \psi_{\bm k}^\alpha \alpha^\dag_{-\bm k}+\sinh \psi_{\bm k}^\beta \beta_{-\bm k}+\cosh \psi_{\bm k}^\beta \beta^\dag_{\bm k})+S\cos\phi, \\
S^{+'}_{B,\bm k}&=\frac{\sqrt{S}}{2}(\cos\phi+1) (-\cosh \psi_{\bm k}^\alpha \alpha_{-\bm k}-\sinh \psi_{\bm k}^\alpha \alpha^\dag_{\bm k}+\cosh \psi_{\bm k}^\beta \beta_{\bm k}+\sinh \psi_{\bm k}^\beta \beta^\dag_{-\bm k}) \nonumber \\
&\hspace{5mm}+\frac{\sqrt{S}}{2}(\cos\phi-1) (-\sinh \psi_{\bm k}^\alpha \alpha_{-\bm k}-\cosh \psi_{\bm k}^\alpha \alpha^\dag_{\bm k}+\sinh \psi_{\bm k}^\beta \beta_{\bm k}+\cosh \psi_{\bm k}^\beta \beta^\dag_{-\bm k})+ S\sin\phi ,  \\
S^{z'}_{B,\bm k}&=-\frac{\sqrt{S}\sin\phi}{2} (-\cosh \psi_{\bm k}^\alpha \alpha_{-\bm k}-\sinh \psi_{\bm k}^\alpha \alpha^\dag_{\bm k}+\cosh \psi_{\bm k}^\beta \beta_{\bm k}+\sinh \psi_{\bm k}^\beta \beta^\dag_{-\bm k}) \nonumber \\
&\hspace{5mm}-\frac{\sqrt{S}\sin\phi}{2
} (-\sinh \psi_{\bm k}^\alpha \alpha_{-\bm k}-\cosh \psi_{\bm k}^\alpha \alpha^\dag_{\bm k}+\sinh \psi_{\bm k}^\beta \beta_{\bm k}+\cosh \psi_{\bm k}^\beta \beta^\dag_{-\bm k})+S \cos\phi ,
\end{align}
and $S_{\nu,{\bm k}}^{-'} = (S_{\nu,-{\bm k}}^{+'})^\dagger$ ($\nu={\rm A},{\rm B}$).
Using these operators, the spin-spin correlation functions are calculated as 
\begin{align}
\chi^{z'z'}_{{\bm k},-{\bm k}}(\tau)=&\frac{S \sin^2\phi}{2}\Bigl[  (K^+_{\bm k \alpha})^2 \Bigl\{ G_{\bm k \alpha}(\tau)+G_{\bm k \alpha}(-\tau) \Bigr\} +  (K^+_{\bm k \beta})^2 \Bigl\{G_{\bm k \beta}(\tau)+G_{\bm k \beta}(-\tau) \Bigr\} \Bigr] , \\
\chi^{+'+'}_{{\bm k},{-\bm k}}(\tau)=
&\frac{S}{2} \Bigl\{ (K^+_{\bm k \alpha})^2 \cos^2 \phi - (K^-_{\bm k \alpha})^2 \Bigr\} \Bigl[  G_{\bm k \alpha}(\tau)+G_{\bm k \alpha}(-\tau) \Bigr] \nonumber \\
&\hspace{2mm}+\frac{S}{2}\Bigl\{ (K^+_{\bm k \beta})^2 \cos^2 \phi - (K^-_{\bm k \beta})^2 \Bigr\} \Bigl[  G_{\bm k \beta}(\tau)+G_{\bm k \beta}(-\tau) \Bigr], \\
\chi^{+'-'}_{{\bm k},{-\bm k}}(\tau)=&\frac{S}{2} \Bigl[ \Bigl\{ K^+_{\bm k \alpha} \cos \phi +  K^-_{\bm k \alpha} \Bigr\}^2 G_{\bm k \alpha}(\tau) +\Bigl\{ K^+_{\bm k \beta} \cos \phi +  K^-_{\bm k \beta} \Bigr\}^2 G_{\bm k \beta}(\tau)  \Bigr] \nonumber \\
&\hspace{2mm}+\frac{S}{2} \Bigl[ \Bigl\{ K^+_{\bm k \alpha} \cos \phi -  K^-_{\bm k \alpha} \Bigr\}^2 G_{\bm k \alpha}(-\tau) +\Bigl\{ K^+_{\bm k \beta} \cos \phi -  K^-_{\bm k \beta} \Bigr\}^2 G_{\bm k \beta}(-\tau)  \Bigr] ,\\
\chi^{-'+'}_{{\bm k},{-\bm k}}(\tau)=&\frac{S}{2} \Bigl[ \Bigl\{ K^+_{\bm k \alpha} \cos \phi -  K^-_{\bm k \alpha} \Bigr\}^2 G_{\bm k \alpha}(\tau) +\Bigl\{ K^+_{\bm k \beta} \cos \phi -  K^-_{\bm k \beta} \Bigr\}^2 G_{\bm k \beta}(\tau)  \Bigr] \nonumber \\
&\hspace{2mm}+\frac{S}{2} \Bigl[ \Bigl\{ K^+_{\bm k \alpha} \cos \phi +  K^-_{\bm k \alpha} \Bigr\}^2 G_{\bm k \alpha}(-\tau) +\Bigl\{ K^+_{\bm k \beta} \cos \phi +  K^-_{\bm k \beta} \Bigr\}^2 G_{\bm k \beta}(-\tau)  \Bigr], 
\end{align}
and $\chi^{-'-'}_{{\bm k},{-\bm k}}(\tau)=\chi^{+'+'}_{{\bm k},{-\bm k}}(\tau)$, where $\cosh \psi_{\bm k}^\nu \pm \sinh \psi_{\bm k}^\nu=K^{\pm}_{{\bm k}\nu}$ and $G_{{\bm k}\nu}(\tau)$ are propagators of the magnons defined by
\begin{align}
G_{{\bm k}\nu}(\tau) = - \frac{1}{\hbar} \langle \nu_{{\bm k}}(\tau) \nu^\dagger_{{\bm k}}(0) \rangle.
\end{align}
Using the diagonalized Hamiltonian, Eq.~(\ref{eq:cant_diag}), the Fourier transformation of this propagator is calculated as
\begin{align}
G_{{\bm k}\nu}(i\omega_n) = \int_0^{\hbar \beta} d\tau\, e^{i\omega_n \tau} G_{{\bm k}\nu}(\tau)
= \frac{1}{i\hbar \omega_n - \varepsilon_{{\bm k}\nu}} .
\end{align}
Combining these results with Eq.~(\ref{GDexpression}), it is straightforward to derive the final result for $\omega_{\bm 0}>0$, which is given in Eqs.~(\ref{canted_result1})-(\ref{canted_result3}).
\end{widetext}

\bibliography{./reference}

\begin{thebibliography}{74}%
\makeatletter
\providecommand \@ifxundefined [1]{%
 \@ifx{#1\undefined}
}%
\providecommand \@ifnum [1]{%
 \ifnum #1\expandafter \@firstoftwo
 \else \expandafter \@secondoftwo
 \fi
}%
\providecommand \@ifx [1]{%
 \ifx #1\expandafter \@firstoftwo
 \else \expandafter \@secondoftwo
 \fi
}%
\providecommand \natexlab [1]{#1}%
\providecommand \enquote  [1]{``#1''}%
\providecommand \bibnamefont  [1]{#1}%
\providecommand \bibfnamefont [1]{#1}%
\providecommand \citenamefont [1]{#1}%
\providecommand \href@noop [0]{\@secondoftwo}%
\providecommand \href [0]{\begingroup \@sanitize@url \@href}%
\providecommand \@href[1]{\@@startlink{#1}\@@href}%
\providecommand \@@href[1]{\endgroup#1\@@endlink}%
\providecommand \@sanitize@url [0]{\catcode `\\12\catcode `\$12\catcode
  `\&12\catcode `\#12\catcode `\^12\catcode `\_12\catcode `\%12\relax}%
\providecommand \@@startlink[1]{}%
\providecommand \@@endlink[0]{}%
\providecommand \url  [0]{\begingroup\@sanitize@url \@url }%
\providecommand \@url [1]{\endgroup\@href {#1}{\urlprefix }}%
\providecommand \urlprefix  [0]{URL }%
\providecommand \Eprint [0]{\href }%
\providecommand \doibase [0]{https://doi.org/}%
\providecommand \selectlanguage [0]{\@gobble}%
\providecommand \bibinfo  [0]{\@secondoftwo}%
\providecommand \bibfield  [0]{\@secondoftwo}%
\providecommand \translation [1]{[#1]}%
\providecommand \BibitemOpen [0]{}%
\providecommand \bibitemStop [0]{}%
\providecommand \bibitemNoStop [0]{.\EOS\space}%
\providecommand \EOS [0]{\spacefactor3000\relax}%
\providecommand \BibitemShut  [1]{\csname bibitem#1\endcsname}%
\let\auto@bib@innerbib\@empty
\bibitem [{\citenamefont {Tserkovnyak}\ \emph {et~al.}(2002)\citenamefont
  {Tserkovnyak}, \citenamefont {Brataas},\ and\ \citenamefont
  {Bauer}}]{Tserkovnyak2002}%
  \BibitemOpen
  \bibfield  {author} {\bibinfo {author} {\bibfnamefont {Y.}~\bibnamefont
  {Tserkovnyak}}, \bibinfo {author} {\bibfnamefont {A.}~\bibnamefont
  {Brataas}},\ and\ \bibinfo {author} {\bibfnamefont {G.~E.~W.}\ \bibnamefont
  {Bauer}},\ }\bibfield  {title} {\bibinfo {title} {{Enhanced Gilbert Damping
  in Thin Ferromagnetic Films}},\ }\href
  {https://doi.org/10.1103/PhysRevLett.88.117601} {\bibfield  {journal}
  {\bibinfo  {journal} {Phys. Rev. Lett.}\ }\textbf {\bibinfo {volume} {88}},\
  \bibinfo {pages} {117601} (\bibinfo {year} {2002})}\BibitemShut {NoStop}%
\bibitem [{\citenamefont {\ifmmode~\check{S}\else \v{S}\fi{}im\'anek}\ and\
  \citenamefont {Heinrich}(2003)}]{Simanek2003}%
  \BibitemOpen
  \bibfield  {author} {\bibinfo {author} {\bibfnamefont {E.}~\bibnamefont
  {\ifmmode~\check{S}\else \v{S}\fi{}im\'anek}}\ and\ \bibinfo {author}
  {\bibfnamefont {B.}~\bibnamefont {Heinrich}},\ }\bibfield  {title} {\bibinfo
  {title} {Gilbert damping in magnetic multilayers},\ }\href
  {https://doi.org/10.1103/PhysRevB.67.144418} {\bibfield  {journal} {\bibinfo
  {journal} {Phys. Rev. B}\ }\textbf {\bibinfo {volume} {67}},\ \bibinfo
  {pages} {144418} (\bibinfo {year} {2003})}\BibitemShut {NoStop}%
\bibitem [{\citenamefont {\ifmmode \check{Z}\else
  \v{Z}\fi{}uti\ifmmode~\acute{c}\else \'{c}\fi{}}\ \emph
  {et~al.}(2004)\citenamefont {\ifmmode \check{Z}\else
  \v{Z}\fi{}uti\ifmmode~\acute{c}\else \'{c}\fi{}}, \citenamefont {Fabian},\
  and\ \citenamefont {Das~Sarma}}]{Zutic2004}%
  \BibitemOpen
  \bibfield  {author} {\bibinfo {author} {\bibfnamefont {I.}~\bibnamefont
  {\ifmmode \check{Z}\else \v{Z}\fi{}uti\ifmmode~\acute{c}\else \'{c}\fi{}}},
  \bibinfo {author} {\bibfnamefont {J.}~\bibnamefont {Fabian}},\ and\ \bibinfo
  {author} {\bibfnamefont {S.}~\bibnamefont {Das~Sarma}},\ }\bibfield  {title}
  {\bibinfo {title} {{Spintronics: Fundamentals and applications}},\ }\href
  {https://doi.org/10.1103/RevModPhys.76.323} {\bibfield  {journal} {\bibinfo
  {journal} {Rev. Mod. Phys.}\ }\textbf {\bibinfo {volume} {76}},\ \bibinfo
  {pages} {323} (\bibinfo {year} {2004})}\BibitemShut {NoStop}%
\bibitem [{\citenamefont {Tserkovnyak}\ \emph {et~al.}(2005)\citenamefont
  {Tserkovnyak}, \citenamefont {Brataas}, \citenamefont {Bauer},\ and\
  \citenamefont {Halperin}}]{Tserkovnyak2005}%
  \BibitemOpen
  \bibfield  {author} {\bibinfo {author} {\bibfnamefont {Y.}~\bibnamefont
  {Tserkovnyak}}, \bibinfo {author} {\bibfnamefont {A.}~\bibnamefont
  {Brataas}}, \bibinfo {author} {\bibfnamefont {G.~E.~W.}\ \bibnamefont
  {Bauer}},\ and\ \bibinfo {author} {\bibfnamefont {B.~I.}\ \bibnamefont
  {Halperin}},\ }\bibfield  {title} {\bibinfo {title} {Nonlocal magnetization
  dynamics in ferromagnetic heterostructures},\ }\href
  {https://doi.org/10.1103/RevModPhys.77.1375} {\bibfield  {journal} {\bibinfo
  {journal} {Rev. Mod. Phys.}\ }\textbf {\bibinfo {volume} {77}},\ \bibinfo
  {pages} {1375} (\bibinfo {year} {2005})}\BibitemShut {NoStop}%
\bibitem [{\citenamefont {Hellman}\ \emph {et~al.}(2017)\citenamefont
  {Hellman}, \citenamefont {Hoffmann}, \citenamefont {Tserkovnyak},
  \citenamefont {Beach}, \citenamefont {Fullerton}, \citenamefont {Leighton},
  \citenamefont {MacDonald}, \citenamefont {Ralph}, \citenamefont {Arena},
  \citenamefont {D\"urr}, \citenamefont {Fischer}, \citenamefont {Grollier},
  \citenamefont {Heremans}, \citenamefont {Jungwirth}, \citenamefont {Kimel},
  \citenamefont {Koopmans}, \citenamefont {Krivorotov}, \citenamefont {May},
  \citenamefont {Petford-Long}, \citenamefont {Rondinelli}, \citenamefont
  {Samarth}, \citenamefont {Schuller}, \citenamefont {Slavin}, \citenamefont
  {Stiles}, \citenamefont {Tchernyshyov}, \citenamefont {Thiaville},\ and\
  \citenamefont {Zink}}]{Hellman2017}%
  \BibitemOpen
  \bibfield  {author} {\bibinfo {author} {\bibfnamefont {F.}~\bibnamefont
  {Hellman}}, \bibinfo {author} {\bibfnamefont {A.}~\bibnamefont {Hoffmann}},
  \bibinfo {author} {\bibfnamefont {Y.}~\bibnamefont {Tserkovnyak}}, \bibinfo
  {author} {\bibfnamefont {G.~S.~D.}\ \bibnamefont {Beach}}, \bibinfo {author}
  {\bibfnamefont {E.~E.}\ \bibnamefont {Fullerton}}, \bibinfo {author}
  {\bibfnamefont {C.}~\bibnamefont {Leighton}}, \bibinfo {author}
  {\bibfnamefont {A.~H.}\ \bibnamefont {MacDonald}}, \bibinfo {author}
  {\bibfnamefont {D.~C.}\ \bibnamefont {Ralph}}, \bibinfo {author}
  {\bibfnamefont {D.~A.}\ \bibnamefont {Arena}}, \bibinfo {author}
  {\bibfnamefont {H.~A.}\ \bibnamefont {D\"urr}}, \bibinfo {author}
  {\bibfnamefont {P.}~\bibnamefont {Fischer}}, \bibinfo {author} {\bibfnamefont
  {J.}~\bibnamefont {Grollier}}, \bibinfo {author} {\bibfnamefont {J.~P.}\
  \bibnamefont {Heremans}}, \bibinfo {author} {\bibfnamefont {T.}~\bibnamefont
  {Jungwirth}}, \bibinfo {author} {\bibfnamefont {A.~V.}\ \bibnamefont
  {Kimel}}, \bibinfo {author} {\bibfnamefont {B.}~\bibnamefont {Koopmans}},
  \bibinfo {author} {\bibfnamefont {I.~N.}\ \bibnamefont {Krivorotov}},
  \bibinfo {author} {\bibfnamefont {S.~J.}\ \bibnamefont {May}}, \bibinfo
  {author} {\bibfnamefont {A.~K.}\ \bibnamefont {Petford-Long}}, \bibinfo
  {author} {\bibfnamefont {J.~M.}\ \bibnamefont {Rondinelli}}, \bibinfo
  {author} {\bibfnamefont {N.}~\bibnamefont {Samarth}}, \bibinfo {author}
  {\bibfnamefont {I.~K.}\ \bibnamefont {Schuller}}, \bibinfo {author}
  {\bibfnamefont {A.~N.}\ \bibnamefont {Slavin}}, \bibinfo {author}
  {\bibfnamefont {M.~D.}\ \bibnamefont {Stiles}}, \bibinfo {author}
  {\bibfnamefont {O.}~\bibnamefont {Tchernyshyov}}, \bibinfo {author}
  {\bibfnamefont {A.}~\bibnamefont {Thiaville}},\ and\ \bibinfo {author}
  {\bibfnamefont {B.~L.}\ \bibnamefont {Zink}},\ }\bibfield  {title} {\bibinfo
  {title} {Interface-induced phenomena in magnetism},\ }\href
  {https://doi.org/10.1103/RevModPhys.89.025006} {\bibfield  {journal}
  {\bibinfo  {journal} {Rev. Mod. Phys.}\ }\textbf {\bibinfo {volume} {89}},\
  \bibinfo {pages} {025006} (\bibinfo {year} {2017})}\BibitemShut {NoStop}%
\bibitem [{\citenamefont {Tsymbal}\ and\ \citenamefont
  {Zuti{\'c}}(2019)}]{Tsymbal2019}%
  \BibitemOpen
  \bibinfo {editor} {\bibfnamefont {E.~Y.}\ \bibnamefont {Tsymbal}}\ and\
  \bibinfo {editor} {\bibfnamefont {I.}~\bibnamefont {Zuti{\'c}}},\ eds.,\
  \href@noop {} {\emph {\bibinfo {title} {Spintronics Handbook, Second Edition:
  Spin Transport and Magnetism}}}\ (\bibinfo  {publisher} {CRC Press},\
  \bibinfo {year} {2019})\BibitemShut {NoStop}%
\bibitem [{\citenamefont {Kamra}\ and\ \citenamefont
  {Belzig}(2016)}]{Kamra2016}%
  \BibitemOpen
  \bibfield  {author} {\bibinfo {author} {\bibfnamefont {A.}~\bibnamefont
  {Kamra}}\ and\ \bibinfo {author} {\bibfnamefont {W.}~\bibnamefont {Belzig}},\
  }\bibfield  {title} {\bibinfo {title} {Super-poissonian shot noise of
  squeezed-magnon mediated spin transport},\ }\href
  {https://doi.org/10.1103/PhysRevLett.116.146601} {\bibfield  {journal}
  {\bibinfo  {journal} {Phys. Rev. Lett.}\ }\textbf {\bibinfo {volume} {116}},\
  \bibinfo {pages} {146601} (\bibinfo {year} {2016})}\BibitemShut {NoStop}%
\bibitem [{\citenamefont {Kamra}\ and\ \citenamefont
  {Belzig}(2017)}]{Kamra2017}%
  \BibitemOpen
  \bibfield  {author} {\bibinfo {author} {\bibfnamefont {A.}~\bibnamefont
  {Kamra}}\ and\ \bibinfo {author} {\bibfnamefont {W.}~\bibnamefont {Belzig}},\
  }\bibfield  {title} {\bibinfo {title} {Spin pumping and shot noise in
  ferrimagnets: Bridging ferro- and antiferromagnets},\ }\href
  {https://doi.org/10.1103/PhysRevLett.119.197201} {\bibfield  {journal}
  {\bibinfo  {journal} {Phys. Rev. Lett.}\ }\textbf {\bibinfo {volume} {119}},\
  \bibinfo {pages} {197201} (\bibinfo {year} {2017})}\BibitemShut {NoStop}%
\bibitem [{\citenamefont {Han}\ \emph {et~al.}(2020)\citenamefont {Han},
  \citenamefont {Maekawa},\ and\ \citenamefont {Xie}}]{Han2020}%
  \BibitemOpen
  \bibfield  {author} {\bibinfo {author} {\bibfnamefont {W.}~\bibnamefont
  {Han}}, \bibinfo {author} {\bibfnamefont {S.}~\bibnamefont {Maekawa}},\ and\
  \bibinfo {author} {\bibfnamefont {X.}~\bibnamefont {Xie}},\ }\bibfield
  {title} {\bibinfo {title} {Spin current as a probe of quantum materials},\
  }\href {https://doi.org/10.1038/s41563-019-0456-7} {\bibfield  {journal}
  {\bibinfo  {journal} {Nat. Mater.}\ }\textbf {\bibinfo {volume} {19}},\
  \bibinfo {pages} {139–152} (\bibinfo {year} {2020})}\BibitemShut {NoStop}%
\bibitem [{\citenamefont {Yang}\ and\ \citenamefont {Hammel}(2018)}]{Yang2018}%
  \BibitemOpen
  \bibfield  {author} {\bibinfo {author} {\bibfnamefont {F.}~\bibnamefont
  {Yang}}\ and\ \bibinfo {author} {\bibfnamefont {P.~C.}\ \bibnamefont
  {Hammel}},\ }\bibfield  {title} {\bibinfo {title} {{FMR-driven spin pumping
  in ${\rm Y}_3{\rm Fe}_5{\rm O}_{12}$-based structures}},\ }\href
  {https://doi.org/10.1088/1361-6463/aac249} {\bibfield  {journal} {\bibinfo
  {journal} {J. Phys. D Appl. Phys.}\ }\textbf {\bibinfo {volume} {51}},\
  \bibinfo {pages} {253001} (\bibinfo {year} {2018})}\BibitemShut {NoStop}%
\bibitem [{\citenamefont {R\'ozsa}\ \emph {et~al.}(2018)\citenamefont
  {R\'ozsa}, \citenamefont {Hagemeister}, \citenamefont {Vedmedenko},\ and\
  \citenamefont {Wiesendanger}}]{Rozsa2018}%
  \BibitemOpen
  \bibfield  {author} {\bibinfo {author} {\bibfnamefont {L.}~\bibnamefont
  {R\'ozsa}}, \bibinfo {author} {\bibfnamefont {J.}~\bibnamefont
  {Hagemeister}}, \bibinfo {author} {\bibfnamefont {E.~Y.}\ \bibnamefont
  {Vedmedenko}},\ and\ \bibinfo {author} {\bibfnamefont {R.}~\bibnamefont
  {Wiesendanger}},\ }\bibfield  {title} {\bibinfo {title} {Effective damping
  enhancement in noncollinear spin structures},\ }\href
  {https://doi.org/10.1103/PhysRevB.98.100404} {\bibfield  {journal} {\bibinfo
  {journal} {Phys. Rev. B}\ }\textbf {\bibinfo {volume} {98}},\ \bibinfo
  {pages} {100404} (\bibinfo {year} {2018})}\BibitemShut {NoStop}%
\bibitem [{\citenamefont {Qiu}\ \emph {et~al.}(2016)\citenamefont {Qiu},
  \citenamefont {Li}, \citenamefont {Hou}, \citenamefont {Arenholz},
  \citenamefont {N'Diaye}, \citenamefont {Tan}, \citenamefont {Uchida},
  \citenamefont {Sato}, \citenamefont {Okamoto}, \citenamefont {Tserkovnyak},
  \citenamefont {Qiu},\ and\ \citenamefont {Saitoh}}]{Qiu2016}%
  \BibitemOpen
  \bibfield  {author} {\bibinfo {author} {\bibfnamefont {Z.}~\bibnamefont
  {Qiu}}, \bibinfo {author} {\bibfnamefont {J.}~\bibnamefont {Li}}, \bibinfo
  {author} {\bibfnamefont {D.}~\bibnamefont {Hou}}, \bibinfo {author}
  {\bibfnamefont {E.}~\bibnamefont {Arenholz}}, \bibinfo {author}
  {\bibfnamefont {A.~T.}\ \bibnamefont {N'Diaye}}, \bibinfo {author}
  {\bibfnamefont {A.}~\bibnamefont {Tan}}, \bibinfo {author} {\bibfnamefont
  {K.-i.}\ \bibnamefont {Uchida}}, \bibinfo {author} {\bibfnamefont
  {K.}~\bibnamefont {Sato}}, \bibinfo {author} {\bibfnamefont {S.}~\bibnamefont
  {Okamoto}}, \bibinfo {author} {\bibfnamefont {Y.}~\bibnamefont
  {Tserkovnyak}}, \bibinfo {author} {\bibfnamefont {Z.~Q.}\ \bibnamefont
  {Qiu}},\ and\ \bibinfo {author} {\bibfnamefont {E.}~\bibnamefont {Saitoh}},\
  }\bibfield  {title} {\bibinfo {title} {Spin-current probe for phase
  transition in an insulator},\ }\href@noop {} {\bibfield  {journal} {\bibinfo
  {journal} {Nat. Commun.}\ }\textbf {\bibinfo {volume} {7}},\ \bibinfo {pages}
  {1} (\bibinfo {year} {2016})}\BibitemShut {NoStop}%
\bibitem [{\citenamefont {Yamamoto}\ \emph {et~al.}(2021)\citenamefont
  {Yamamoto}, \citenamefont {Kato},\ and\ \citenamefont
  {Matsuo}}]{Yamamoto2021}%
  \BibitemOpen
  \bibfield  {author} {\bibinfo {author} {\bibfnamefont {T.}~\bibnamefont
  {Yamamoto}}, \bibinfo {author} {\bibfnamefont {T.}~\bibnamefont {Kato}},\
  and\ \bibinfo {author} {\bibfnamefont {M.}~\bibnamefont {Matsuo}},\
  }\bibfield  {title} {\bibinfo {title} {{Spin current at a magnetic junction
  as a probe of the Kondo state}},\ }\href
  {https://doi.org/10.1103/PhysRevB.104.L121401} {\bibfield  {journal}
  {\bibinfo  {journal} {Phys. Rev. B}\ }\textbf {\bibinfo {volume} {104}},\
  \bibinfo {pages} {L121401} (\bibinfo {year} {2021})}\BibitemShut {NoStop}%
\bibitem [{\citenamefont {Ominato}\ and\ \citenamefont
  {Matsuo}(2020)}]{Ominato2020a}%
  \BibitemOpen
  \bibfield  {author} {\bibinfo {author} {\bibfnamefont {Y.}~\bibnamefont
  {Ominato}}\ and\ \bibinfo {author} {\bibfnamefont {M.}~\bibnamefont
  {Matsuo}},\ }\bibfield  {title} {\bibinfo {title} {{Quantum Oscillations of
  Gilbert Damping in Ferromagnetic/Graphene Bilayer Systems}},\ }\href
  {https://doi.org/https://doi.org/10.7566/JPSJ.89.053704} {\bibfield
  {journal} {\bibinfo  {journal} {J. Phys. Soc. Jpn.}\ }\textbf {\bibinfo
  {volume} {89}},\ \bibinfo {pages} {053704} (\bibinfo {year}
  {2020})}\BibitemShut {NoStop}%
\bibitem [{\citenamefont {Ominato}\ \emph {et~al.}(2020)\citenamefont
  {Ominato}, \citenamefont {Fujimoto},\ and\ \citenamefont
  {Matsuo}}]{Ominato2020b}%
  \BibitemOpen
  \bibfield  {author} {\bibinfo {author} {\bibfnamefont {Y.}~\bibnamefont
  {Ominato}}, \bibinfo {author} {\bibfnamefont {J.}~\bibnamefont {Fujimoto}},\
  and\ \bibinfo {author} {\bibfnamefont {M.}~\bibnamefont {Matsuo}},\
  }\bibfield  {title} {\bibinfo {title} {{Valley-Dependent Spin Transport in
  Monolayer Transition-Metal Dichalcogenides}},\ }\href
  {https://doi.org/10.1103/PhysRevLett.124.166803} {\bibfield  {journal}
  {\bibinfo  {journal} {Phys. Rev. Lett.}\ }\textbf {\bibinfo {volume} {124}},\
  \bibinfo {pages} {166803} (\bibinfo {year} {2020})}\BibitemShut {NoStop}%
\bibitem [{\citenamefont {Yama}\ \emph {et~al.}(2021)\citenamefont {Yama},
  \citenamefont {Tatsuno}, \citenamefont {Kato},\ and\ \citenamefont
  {Matsuo}}]{Yama2021}%
  \BibitemOpen
  \bibfield  {author} {\bibinfo {author} {\bibfnamefont {M.}~\bibnamefont
  {Yama}}, \bibinfo {author} {\bibfnamefont {M.}~\bibnamefont {Tatsuno}},
  \bibinfo {author} {\bibfnamefont {T.}~\bibnamefont {Kato}},\ and\ \bibinfo
  {author} {\bibfnamefont {M.}~\bibnamefont {Matsuo}},\ }\bibfield  {title}
  {\bibinfo {title} {{Spin pumping of two-dimensional electron gas with Rashba
  and Dresselhaus spin-orbit interactions}},\ }\href
  {https://doi.org/10.1103/PhysRevB.104.054410} {\bibfield  {journal} {\bibinfo
   {journal} {Phys. Rev. B}\ }\textbf {\bibinfo {volume} {104}},\ \bibinfo
  {pages} {054410} (\bibinfo {year} {2021})}\BibitemShut {NoStop}%
\bibitem [{\citenamefont {Inoue}\ \emph {et~al.}(2017)\citenamefont {Inoue},
  \citenamefont {Ichioka},\ and\ \citenamefont {Adachi}}]{Inoue2017}%
  \BibitemOpen
  \bibfield  {author} {\bibinfo {author} {\bibfnamefont {M.}~\bibnamefont
  {Inoue}}, \bibinfo {author} {\bibfnamefont {M.}~\bibnamefont {Ichioka}},\
  and\ \bibinfo {author} {\bibfnamefont {H.}~\bibnamefont {Adachi}},\
  }\bibfield  {title} {\bibinfo {title} {Spin pumping into superconductors: A
  new probe of spin dynamics in a superconducting thin film},\ }\href
  {https://doi.org/10.1103/PhysRevB.96.024414} {\bibfield  {journal} {\bibinfo
  {journal} {Phys. Rev. B}\ }\textbf {\bibinfo {volume} {96}},\ \bibinfo
  {pages} {024414} (\bibinfo {year} {2017})}\BibitemShut {NoStop}%
\bibitem [{\citenamefont {Silaev}(2020{\natexlab{a}})}]{Silaev2020}%
  \BibitemOpen
  \bibfield  {author} {\bibinfo {author} {\bibfnamefont {M.~A.}\ \bibnamefont
  {Silaev}},\ }\bibfield  {title} {\bibinfo {title} {{Finite-frequency spin
  susceptibility and spin pumping in superconductors with spin-orbit
  relaxation}},\ }\href {https://doi.org/10.1103/PhysRevB.102.144521}
  {\bibfield  {journal} {\bibinfo  {journal} {Phys. Rev. B}\ }\textbf {\bibinfo
  {volume} {102}},\ \bibinfo {pages} {144521} (\bibinfo {year}
  {2020}{\natexlab{a}})}\BibitemShut {NoStop}%
\bibitem [{\citenamefont {Silaev}(2020{\natexlab{b}})}]{Silaev2020b}%
  \BibitemOpen
  \bibfield  {author} {\bibinfo {author} {\bibfnamefont {M.~A.}\ \bibnamefont
  {Silaev}},\ }\bibfield  {title} {\bibinfo {title} {{Large enhancement of spin
  pumping due to the surface bound states in normal metal--superconductor
  structures}},\ }\href {https://doi.org/10.1103/PhysRevB.102.180502}
  {\bibfield  {journal} {\bibinfo  {journal} {Phys. Rev. B}\ }\textbf {\bibinfo
  {volume} {102}},\ \bibinfo {pages} {180502(R)} (\bibinfo {year}
  {2020}{\natexlab{b}})}\BibitemShut {NoStop}%
\bibitem [{\citenamefont {Simensen}\ \emph {et~al.}(2021)\citenamefont
  {Simensen}, \citenamefont {Johnsen}, \citenamefont {Linder},\ and\
  \citenamefont {Brataas}}]{Simensen2021}%
  \BibitemOpen
  \bibfield  {author} {\bibinfo {author} {\bibfnamefont {H.~T.}\ \bibnamefont
  {Simensen}}, \bibinfo {author} {\bibfnamefont {L.~G.}\ \bibnamefont
  {Johnsen}}, \bibinfo {author} {\bibfnamefont {J.}~\bibnamefont {Linder}},\
  and\ \bibinfo {author} {\bibfnamefont {A.}~\bibnamefont {Brataas}},\
  }\bibfield  {title} {\bibinfo {title} {Spin pumping between noncollinear
  ferromagnetic insulators through thin superconductors},\ }\href
  {https://doi.org/10.1103/PhysRevB.103.024524} {\bibfield  {journal} {\bibinfo
   {journal} {Phys. Rev. B}\ }\textbf {\bibinfo {volume} {103}},\ \bibinfo
  {pages} {024524} (\bibinfo {year} {2021})}\BibitemShut {NoStop}%
\bibitem [{\citenamefont {Fyhn}\ and\ \citenamefont {Linder}(2021)}]{Fyhn2021}%
  \BibitemOpen
  \bibfield  {author} {\bibinfo {author} {\bibfnamefont {E.~H.}\ \bibnamefont
  {Fyhn}}\ and\ \bibinfo {author} {\bibfnamefont {J.}~\bibnamefont {Linder}},\
  }\bibfield  {title} {\bibinfo {title} {Spin pumping in
  superconductor-antiferromagnetic insulator bilayers},\ }\href
  {https://doi.org/10.1103/PhysRevB.103.134508} {\bibfield  {journal} {\bibinfo
   {journal} {Phys. Rev. B}\ }\textbf {\bibinfo {volume} {103}},\ \bibinfo
  {pages} {134508} (\bibinfo {year} {2021})}\BibitemShut {NoStop}%
\bibitem [{\citenamefont {Ominato}\ \emph
  {et~al.}(2022{\natexlab{a}})\citenamefont {Ominato}, \citenamefont
  {Yamakage}, \citenamefont {Kato},\ and\ \citenamefont
  {Matsuo}}]{Ominato2022a}%
  \BibitemOpen
  \bibfield  {author} {\bibinfo {author} {\bibfnamefont {Y.}~\bibnamefont
  {Ominato}}, \bibinfo {author} {\bibfnamefont {A.}~\bibnamefont {Yamakage}},
  \bibinfo {author} {\bibfnamefont {T.}~\bibnamefont {Kato}},\ and\ \bibinfo
  {author} {\bibfnamefont {M.}~\bibnamefont {Matsuo}},\ }\bibfield  {title}
  {\bibinfo {title} {{Ferromagnetic resonance modulation in $d$-wave
  superconductor/ferromagnetic insulator bilayer systems}},\ }\href
  {https://doi.org/10.1103/PhysRevB.105.205406} {\bibfield  {journal} {\bibinfo
   {journal} {Phys. Rev. B}\ }\textbf {\bibinfo {volume} {105}},\ \bibinfo
  {pages} {205406} (\bibinfo {year} {2022}{\natexlab{a}})}\BibitemShut
  {NoStop}%
\bibitem [{\citenamefont {Ominato}\ \emph
  {et~al.}(2022{\natexlab{b}})\citenamefont {Ominato}, \citenamefont
  {Yamakage},\ and\ \citenamefont {Matsuo}}]{Ominato2022b}%
  \BibitemOpen
  \bibfield  {author} {\bibinfo {author} {\bibfnamefont {Y.}~\bibnamefont
  {Ominato}}, \bibinfo {author} {\bibfnamefont {A.}~\bibnamefont {Yamakage}},\
  and\ \bibinfo {author} {\bibfnamefont {M.}~\bibnamefont {Matsuo}},\
  }\bibfield  {title} {\bibinfo {title} {{Anisotropic superconducting spin
  transport at magnetic interfaces}},\ }\href
  {https://doi.org/10.1103/PhysRevB.106.L161406} {\bibfield  {journal}
  {\bibinfo  {journal} {Phys. Rev. B}\ }\textbf {\bibinfo {volume} {106}},\
  \bibinfo {pages} {L161406} (\bibinfo {year}
  {2022}{\natexlab{b}})}\BibitemShut {NoStop}%
\bibitem [{\citenamefont {Funato}\ \emph {et~al.}(2022)\citenamefont {Funato},
  \citenamefont {Kato},\ and\ \citenamefont {Matsuo}}]{Funato2022}%
  \BibitemOpen
  \bibfield  {author} {\bibinfo {author} {\bibfnamefont {T.}~\bibnamefont
  {Funato}}, \bibinfo {author} {\bibfnamefont {T.}~\bibnamefont {Kato}},\ and\
  \bibinfo {author} {\bibfnamefont {M.}~\bibnamefont {Matsuo}},\ }\bibfield
  {title} {\bibinfo {title} {{Spin pumping into anisotropic Dirac electrons}},\
  }\href {https://doi.org/10.1103/PhysRevB.106.144418} {\bibfield  {journal}
  {\bibinfo  {journal} {Phys. Rev. B}\ }\textbf {\bibinfo {volume} {106}},\
  \bibinfo {pages} {144418} (\bibinfo {year} {2022})}\BibitemShut {NoStop}%
\bibitem [{\citenamefont {Sun}\ and\ \citenamefont
  {Linder}(2023{\natexlab{a}})}]{Sun2023}%
  \BibitemOpen
  \bibfield  {author} {\bibinfo {author} {\bibfnamefont {C.}~\bibnamefont
  {Sun}}\ and\ \bibinfo {author} {\bibfnamefont {J.}~\bibnamefont {Linder}},\
  }\bibfield  {title} {\bibinfo {title} {{Spin pumping from a ferromagnetic
  insulator to an unconventional superconductor with interfacial Andreev bound
  states}},\ }\href {https://doi.org/10.1103/PhysRevB.107.144504} {\bibfield
  {journal} {\bibinfo  {journal} {Phys. Rev. B}\ }\textbf {\bibinfo {volume}
  {107}},\ \bibinfo {pages} {144504} (\bibinfo {year}
  {2023}{\natexlab{a}})}\BibitemShut {NoStop}%
\bibitem [{\citenamefont {Funaki}\ \emph {et~al.}(2023)\citenamefont {Funaki},
  \citenamefont {Yamakage},\ and\ \citenamefont {Matsuo}}]{Funaki2023}%
  \BibitemOpen
  \bibfield  {author} {\bibinfo {author} {\bibfnamefont {H.}~\bibnamefont
  {Funaki}}, \bibinfo {author} {\bibfnamefont {A.}~\bibnamefont {Yamakage}},\
  and\ \bibinfo {author} {\bibfnamefont {M.}~\bibnamefont {Matsuo}},\
  }\bibfield  {title} {\bibinfo {title} {{Anisotropic spin-current spectroscopy
  of ferromagnetic superconducting gap symmetries}},\ }\href
  {https://doi.org/10.1103/PhysRevB.107.184437} {\bibfield  {journal} {\bibinfo
   {journal} {Phys. Rev. B}\ }\textbf {\bibinfo {volume} {107}},\ \bibinfo
  {pages} {184437} (\bibinfo {year} {2023})}\BibitemShut {NoStop}%
\bibitem [{\citenamefont {Sun}\ and\ \citenamefont
  {Linder}(2023{\natexlab{b}})}]{Sun2023b}%
  \BibitemOpen
  \bibfield  {author} {\bibinfo {author} {\bibfnamefont {C.}~\bibnamefont
  {Sun}}\ and\ \bibinfo {author} {\bibfnamefont {J.}~\bibnamefont {Linder}},\
  }\bibfield  {title} {\bibinfo {title} {Spin pumping from a ferromagnetic
  insulator into an altermagnet},\ }\href
  {https://doi.org/10.1103/PhysRevB.108.L140408} {\bibfield  {journal}
  {\bibinfo  {journal} {Phys. Rev. B}\ }\textbf {\bibinfo {volume} {108}},\
  \bibinfo {pages} {L140408} (\bibinfo {year}
  {2023}{\natexlab{b}})}\BibitemShut {NoStop}%
\bibitem [{\citenamefont {Yama}\ \emph {et~al.}(2023)\citenamefont {Yama},
  \citenamefont {Matsuo},\ and\ \citenamefont {Kato}}]{Yama2023a}%
  \BibitemOpen
  \bibfield  {author} {\bibinfo {author} {\bibfnamefont {M.}~\bibnamefont
  {Yama}}, \bibinfo {author} {\bibfnamefont {M.}~\bibnamefont {Matsuo}},\ and\
  \bibinfo {author} {\bibfnamefont {T.}~\bibnamefont {Kato}},\ }\bibfield
  {title} {\bibinfo {title} {{Effect of vertex corrections on the enhancement
  of Gilbert damping in spin pumping into a two-dimensional electron gas}},\
  }\href {https://doi.org/10.1103/PhysRevB.107.174414} {\bibfield  {journal}
  {\bibinfo  {journal} {Phys. Rev. B}\ }\textbf {\bibinfo {volume} {107}},\
  \bibinfo {pages} {174414} (\bibinfo {year} {2023})}\BibitemShut {NoStop}%
\bibitem [{\citenamefont {Fukuzawa}\ \emph {et~al.}(2023)\citenamefont
  {Fukuzawa}, \citenamefont {Kato}, \citenamefont {Matsuo}, \citenamefont
  {Jonckheere}, \citenamefont {Rech},\ and\ \citenamefont
  {Martin}}]{Fukuzawa2023}%
  \BibitemOpen
  \bibfield  {author} {\bibinfo {author} {\bibfnamefont {K.}~\bibnamefont
  {Fukuzawa}}, \bibinfo {author} {\bibfnamefont {T.}~\bibnamefont {Kato}},
  \bibinfo {author} {\bibfnamefont {M.}~\bibnamefont {Matsuo}}, \bibinfo
  {author} {\bibfnamefont {T.}~\bibnamefont {Jonckheere}}, \bibinfo {author}
  {\bibfnamefont {J.}~\bibnamefont {Rech}},\ and\ \bibinfo {author}
  {\bibfnamefont {T.}~\bibnamefont {Martin}},\ }\bibfield  {title} {\bibinfo
  {title} {Spin pumping into carbon nanotubes},\ }\href
  {https://doi.org/10.1103/PhysRevB.108.134429} {\bibfield  {journal} {\bibinfo
   {journal} {Phys. Rev. B}\ }\textbf {\bibinfo {volume} {108}},\ \bibinfo
  {pages} {134429} (\bibinfo {year} {2023})}\BibitemShut {NoStop}%
\bibitem [{\citenamefont {Haddad}\ \emph {et~al.}(2023)\citenamefont {Haddad},
  \citenamefont {Kato}, \citenamefont {Zhu},\ and\ \citenamefont
  {Mandhour}}]{Haddad2023}%
  \BibitemOpen
  \bibfield  {author} {\bibinfo {author} {\bibfnamefont {S.}~\bibnamefont
  {Haddad}}, \bibinfo {author} {\bibfnamefont {T.}~\bibnamefont {Kato}},
  \bibinfo {author} {\bibfnamefont {J.}~\bibnamefont {Zhu}},\ and\ \bibinfo
  {author} {\bibfnamefont {L.}~\bibnamefont {Mandhour}},\ }\bibfield  {title}
  {\bibinfo {title} {Twisted bilayer graphene reveals its flat bands under spin
  pumping},\ }\href {https://doi.org/10.1103/PhysRevB.108.L121101} {\bibfield
  {journal} {\bibinfo  {journal} {Phys. Rev. B}\ }\textbf {\bibinfo {volume}
  {108}},\ \bibinfo {pages} {L121101} (\bibinfo {year} {2023})}\BibitemShut
  {NoStop}%
\bibitem [{\citenamefont {Furuya}\ \emph {et~al.}(2024)\citenamefont {Furuya},
  \citenamefont {Matsuo},\ and\ \citenamefont {Kato}}]{Furuya2024}%
  \BibitemOpen
  \bibfield  {author} {\bibinfo {author} {\bibfnamefont {S.~C.}\ \bibnamefont
  {Furuya}}, \bibinfo {author} {\bibfnamefont {M.}~\bibnamefont {Matsuo}},\
  and\ \bibinfo {author} {\bibfnamefont {T.}~\bibnamefont {Kato}},\ }\bibfield
  {title} {\bibinfo {title} {Spin pumping into quantum spin chains},\ }\href
  {https://doi.org/10.1103/PhysRevB.110.165129} {\bibfield  {journal} {\bibinfo
   {journal} {Phys. Rev. B}\ }\textbf {\bibinfo {volume} {110}},\ \bibinfo
  {pages} {165129} (\bibinfo {year} {2024})}\BibitemShut {NoStop}%
\bibitem [{\citenamefont {Yama}\ \emph {et~al.}(2024)\citenamefont {Yama},
  \citenamefont {Matsuo},\ and\ \citenamefont {Kato}}]{Yama2024proceeding}%
  \BibitemOpen
  \bibfield  {author} {\bibinfo {author} {\bibfnamefont {M.}~\bibnamefont
  {Yama}}, \bibinfo {author} {\bibfnamefont {M.}~\bibnamefont {Matsuo}},\ and\
  \bibinfo {author} {\bibfnamefont {T.}~\bibnamefont {Kato}},\ }\bibfield
  {title} {\bibinfo {title} {{Theory of spin pumping and inverse
  Rashba-Edelstein effect in a two-dimensional electron gas}},\ }in\ \href
  {https://doi.org/10.1117/12.3027001} {\emph {\bibinfo {booktitle}
  {Spintronics XVII}}},\ Vol.\ \bibinfo {volume} {13119},\ \bibinfo {editor}
  {edited by\ \bibinfo {editor} {\bibfnamefont {J.-E.}\ \bibnamefont
  {Wegrowe}}, \bibinfo {editor} {\bibfnamefont {J.~S.}\ \bibnamefont
  {Friedman}}, \bibinfo {editor} {\bibfnamefont {M.}~\bibnamefont {Razeghi}},\
  and\ \bibinfo {editor} {\bibfnamefont {H.}~\bibnamefont {Jaffr{\`e}s}}},\
  \bibinfo {organization} {International Society for Optics and Photonics}\
  (\bibinfo  {publisher} {SPIE},\ \bibinfo {year} {2024})\ p.\ \bibinfo {pages}
  {131190J}\BibitemShut {NoStop}%
\bibitem [{\citenamefont {Blume}\ and\ \citenamefont
  {Hsieh}(1969)}]{Blume1969}%
  \BibitemOpen
  \bibfield  {author} {\bibinfo {author} {\bibfnamefont {M.}~\bibnamefont
  {Blume}}\ and\ \bibinfo {author} {\bibfnamefont {Y.~Y.~H.}\ \bibnamefont
  {Hsieh}},\ }\bibfield  {title} {\bibinfo {title} {{Biquadratic Exchange and
  Quadrupolar Ordering}},\ }\href {https://doi.org/10.1063/1.1657616}
  {\bibfield  {journal} {\bibinfo  {journal} {J. Appl. Phys.}\ }\textbf
  {\bibinfo {volume} {40}},\ \bibinfo {pages} {1249} (\bibinfo {year}
  {1969})}\BibitemShut {NoStop}%
\bibitem [{\citenamefont {Chen}\ and\ \citenamefont {Levy}(1971)}]{Chen1971}%
  \BibitemOpen
  \bibfield  {author} {\bibinfo {author} {\bibfnamefont {H.~H.}\ \bibnamefont
  {Chen}}\ and\ \bibinfo {author} {\bibfnamefont {P.~M.}\ \bibnamefont
  {Levy}},\ }\bibfield  {title} {\bibinfo {title} {{Quadrupole Phase
  Transitions in Magnetic Solids}},\ }\href
  {https://doi.org/10.1103/PhysRevLett.27.1383} {\bibfield  {journal} {\bibinfo
   {journal} {Phys. Rev. Lett.}\ }\textbf {\bibinfo {volume} {27}},\ \bibinfo
  {pages} {1383} (\bibinfo {year} {1971})}\BibitemShut {NoStop}%
\bibitem [{\citenamefont {Andreev}\ and\ \citenamefont
  {Grishchuk}(1984)}]{Andreev1984}%
  \BibitemOpen
  \bibfield  {author} {\bibinfo {author} {\bibfnamefont {A.}~\bibnamefont
  {Andreev}}\ and\ \bibinfo {author} {\bibfnamefont {I.}~\bibnamefont
  {Grishchuk}},\ }\bibfield  {title} {\bibinfo {title} {Spin nematics},\
  }\href@noop {} {\bibfield  {journal} {\bibinfo  {journal} {Sov. Phys. JETP}\
  }\textbf {\bibinfo {volume} {60}} (\bibinfo {year} {1984})}\BibitemShut
  {NoStop}%
\bibitem [{\citenamefont {Papanicolaou}(1988)}]{Papanicolaou1988}%
  \BibitemOpen
  \bibfield  {author} {\bibinfo {author} {\bibfnamefont {N.}~\bibnamefont
  {Papanicolaou}},\ }\bibfield  {title} {\bibinfo {title} {Unusual phases in
  quantum spin-1 systems},\ }\href
  {https://doi.org/10.1016/0550-3213(88)90073-9} {\bibfield  {journal}
  {\bibinfo  {journal} {Nucl. Phys. B}\ }\textbf {\bibinfo {volume} {305}},\
  \bibinfo {pages} {367} (\bibinfo {year} {1988})}\BibitemShut {NoStop}%
\bibitem [{\citenamefont {Tanaka}\ \emph {et~al.}(2001)\citenamefont {Tanaka},
  \citenamefont {Tanaka},\ and\ \citenamefont {Idogaki}}]{Tanaka2001}%
  \BibitemOpen
  \bibfield  {author} {\bibinfo {author} {\bibfnamefont {K.}~\bibnamefont
  {Tanaka}}, \bibinfo {author} {\bibfnamefont {A.}~\bibnamefont {Tanaka}},\
  and\ \bibinfo {author} {\bibfnamefont {T.}~\bibnamefont {Idogaki}},\
  }\bibfield  {title} {\bibinfo {title} {{Long-range order in the ground state
  of the $S=1$ isotropic bilinear-biquadratic exchange Hamiltonian}},\ }\href
  {https://doi.org/10.1088/0305-4470/34/42/304} {\bibfield  {journal} {\bibinfo
   {journal} {J. Phys. A Math. Gen.}\ }\textbf {\bibinfo {volume} {34}},\
  \bibinfo {pages} {8767} (\bibinfo {year} {2001})}\BibitemShut {NoStop}%
\bibitem [{\citenamefont {Bhattacharjee}\ \emph {et~al.}(2006)\citenamefont
  {Bhattacharjee}, \citenamefont {Shenoy},\ and\ \citenamefont
  {Senthil}}]{Bhattacharjee2006}%
  \BibitemOpen
  \bibfield  {author} {\bibinfo {author} {\bibfnamefont {S.}~\bibnamefont
  {Bhattacharjee}}, \bibinfo {author} {\bibfnamefont {V.~B.}\ \bibnamefont
  {Shenoy}},\ and\ \bibinfo {author} {\bibfnamefont {T.}~\bibnamefont
  {Senthil}},\ }\bibfield  {title} {\bibinfo {title} {{Possible ferro-spin
  nematic order in $\mathrm{Ni}\mathrm{Ga}_{2}\mathrm{S}_{4}$}},\ }\href
  {https://doi.org/10.1103/PhysRevB.74.092406} {\bibfield  {journal} {\bibinfo
  {journal} {Phys. Rev. B}\ }\textbf {\bibinfo {volume} {74}},\ \bibinfo
  {pages} {092406} (\bibinfo {year} {2006})}\BibitemShut {NoStop}%
\bibitem [{\citenamefont {L\"auchli}\ \emph {et~al.}(2006)\citenamefont
  {L\"auchli}, \citenamefont {Mila},\ and\ \citenamefont {Penc}}]{Lauchli2006}%
  \BibitemOpen
  \bibfield  {author} {\bibinfo {author} {\bibfnamefont {A.}~\bibnamefont
  {L\"auchli}}, \bibinfo {author} {\bibfnamefont {F.}~\bibnamefont {Mila}},\
  and\ \bibinfo {author} {\bibfnamefont {K.}~\bibnamefont {Penc}},\ }\bibfield
  {title} {\bibinfo {title} {{Quadrupolar Phases of the $S=1$
  Bilinear-Biquadratic Heisenberg Model on the Triangular Lattice}},\ }\href
  {https://doi.org/10.1103/PhysRevLett.97.087205} {\bibfield  {journal}
  {\bibinfo  {journal} {Phys. Rev. Lett.}\ }\textbf {\bibinfo {volume} {97}},\
  \bibinfo {pages} {087205} (\bibinfo {year} {2006})}\BibitemShut {NoStop}%
\bibitem [{\citenamefont {Tsunetsugu}\ and\ \citenamefont
  {Arikawa}(2006)}]{Tsunetsugu2006}%
  \BibitemOpen
  \bibfield  {author} {\bibinfo {author} {\bibfnamefont {H.}~\bibnamefont
  {Tsunetsugu}}\ and\ \bibinfo {author} {\bibfnamefont {M.}~\bibnamefont
  {Arikawa}},\ }\bibfield  {title} {\bibinfo {title} {{Spin Nematic Phase in
  $S=1$ Triangular Antiferromagnets}},\ }\href
  {https://doi.org/10.1143/jpsj.75.083701} {\bibfield  {journal} {\bibinfo
  {journal} {J. Phys. Soc. Jpn.}\ }\textbf {\bibinfo {volume} {75}},\ \bibinfo
  {pages} {083701} (\bibinfo {year} {2006})}\BibitemShut {NoStop}%
\bibitem [{\citenamefont {Li}\ \emph {et~al.}(2007)\citenamefont {Li},
  \citenamefont {Zhang},\ and\ \citenamefont {Shen}}]{Li2007}%
  \BibitemOpen
  \bibfield  {author} {\bibinfo {author} {\bibfnamefont {P.}~\bibnamefont
  {Li}}, \bibinfo {author} {\bibfnamefont {G.-M.}\ \bibnamefont {Zhang}},\ and\
  \bibinfo {author} {\bibfnamefont {S.-Q.}\ \bibnamefont {Shen}},\ }\bibfield
  {title} {\bibinfo {title} {{SU(3) bosons and the spin nematic state on the
  spin-1 bilinear-biquadratic triangular lattice}},\ }\href
  {https://doi.org/10.1103/PhysRevB.75.104420} {\bibfield  {journal} {\bibinfo
  {journal} {Phys. Rev. B}\ }\textbf {\bibinfo {volume} {75}},\ \bibinfo
  {pages} {104420} (\bibinfo {year} {2007})}\BibitemShut {NoStop}%
\bibitem [{\citenamefont {Tsunetsugu}\ and\ \citenamefont
  {Arikawa}(2007)}]{Tsunetsugu2007}%
  \BibitemOpen
  \bibfield  {author} {\bibinfo {author} {\bibfnamefont {H.}~\bibnamefont
  {Tsunetsugu}}\ and\ \bibinfo {author} {\bibfnamefont {M.}~\bibnamefont
  {Arikawa}},\ }\bibfield  {title} {\bibinfo {title} {{The spin nematic state
  in triangular antiferromagnets}},\ }\href
  {https://doi.org/10.1088/0953-8984/19/14/145248} {\bibfield  {journal}
  {\bibinfo  {journal} {J. Condens. Matter Phys.}\ }\textbf {\bibinfo {volume}
  {19}},\ \bibinfo {pages} {145248} (\bibinfo {year} {2007})}\BibitemShut
  {NoStop}%
\bibitem [{\citenamefont {Chubukov}(1991)}]{Chubukov1991}%
  \BibitemOpen
  \bibfield  {author} {\bibinfo {author} {\bibfnamefont {A.~V.}\ \bibnamefont
  {Chubukov}},\ }\bibfield  {title} {\bibinfo {title} {Chiral, nematic, and
  dimer states in quantum spin chains},\ }\href
  {https://doi.org/10.1103/PhysRevB.44.4693} {\bibfield  {journal} {\bibinfo
  {journal} {Phys. Rev. B}\ }\textbf {\bibinfo {volume} {44}},\ \bibinfo
  {pages} {4693} (\bibinfo {year} {1991})}\BibitemShut {NoStop}%
\bibitem [{\citenamefont {Shannon}\ \emph {et~al.}(2006)\citenamefont
  {Shannon}, \citenamefont {Momoi},\ and\ \citenamefont
  {Sindzingre}}]{Shannon2006}%
  \BibitemOpen
  \bibfield  {author} {\bibinfo {author} {\bibfnamefont {N.}~\bibnamefont
  {Shannon}}, \bibinfo {author} {\bibfnamefont {T.}~\bibnamefont {Momoi}},\
  and\ \bibinfo {author} {\bibfnamefont {P.}~\bibnamefont {Sindzingre}},\
  }\bibfield  {title} {\bibinfo {title} {{Nematic Order in Square Lattice
  Frustrated Ferromagnets}},\ }\href
  {https://doi.org/10.1103/PhysRevLett.96.027213} {\bibfield  {journal}
  {\bibinfo  {journal} {Phys. Rev. Lett.}\ }\textbf {\bibinfo {volume} {96}},\
  \bibinfo {pages} {027213} (\bibinfo {year} {2006})}\BibitemShut {NoStop}%
\bibitem [{\citenamefont {Ueda}\ and\ \citenamefont
  {Totsuka}(2007)}]{Ueda2007}%
  \BibitemOpen
  \bibfield  {author} {\bibinfo {author} {\bibfnamefont {H.~T.}\ \bibnamefont
  {Ueda}}\ and\ \bibinfo {author} {\bibfnamefont {K.}~\bibnamefont {Totsuka}},\
  }\bibfield  {title} {\bibinfo {title} {{Ground-state phase diagram and
  magnetic properties of a tetramerized spin-$1/2$
  ${J}_{1}\text{\ensuremath{-}}{J}_{2}$ model: BEC of bound magnons and absence
  of the transverse magnetization}},\ }\href
  {https://doi.org/10.1103/PhysRevB.76.214428} {\bibfield  {journal} {\bibinfo
  {journal} {Phys. Rev. B}\ }\textbf {\bibinfo {volume} {76}},\ \bibinfo
  {pages} {214428} (\bibinfo {year} {2007})}\BibitemShut {NoStop}%
\bibitem [{\citenamefont {Vekua}\ \emph {et~al.}(2007)\citenamefont {Vekua},
  \citenamefont {Honecker}, \citenamefont {Mikeska},\ and\ \citenamefont
  {Heidrich-Meisner}}]{Vekua2007}%
  \BibitemOpen
  \bibfield  {author} {\bibinfo {author} {\bibfnamefont {T.}~\bibnamefont
  {Vekua}}, \bibinfo {author} {\bibfnamefont {A.}~\bibnamefont {Honecker}},
  \bibinfo {author} {\bibfnamefont {H.-J.}\ \bibnamefont {Mikeska}},\ and\
  \bibinfo {author} {\bibfnamefont {F.}~\bibnamefont {Heidrich-Meisner}},\
  }\bibfield  {title} {\bibinfo {title} {{Correlation functions and excitation
  spectrum of the frustrated ferromagnetic spin-$1/2$ chain in an external
  magnetic field}},\ }\href {https://doi.org/10.1103/PhysRevB.76.174420}
  {\bibfield  {journal} {\bibinfo  {journal} {Phys. Rev. B}\ }\textbf {\bibinfo
  {volume} {76}},\ \bibinfo {pages} {174420} (\bibinfo {year}
  {2007})}\BibitemShut {NoStop}%
\bibitem [{\citenamefont {Hikihara}\ \emph {et~al.}(2008)\citenamefont
  {Hikihara}, \citenamefont {Kecke}, \citenamefont {Momoi},\ and\ \citenamefont
  {Furusaki}}]{Hikihara2008}%
  \BibitemOpen
  \bibfield  {author} {\bibinfo {author} {\bibfnamefont {T.}~\bibnamefont
  {Hikihara}}, \bibinfo {author} {\bibfnamefont {L.}~\bibnamefont {Kecke}},
  \bibinfo {author} {\bibfnamefont {T.}~\bibnamefont {Momoi}},\ and\ \bibinfo
  {author} {\bibfnamefont {A.}~\bibnamefont {Furusaki}},\ }\bibfield  {title}
  {\bibinfo {title} {{Vector chiral and multipolar orders in the spin-$1/2$
  frustrated ferromagnetic chain in magnetic field}},\ }\href
  {https://doi.org/10.1103/PhysRevB.78.144404} {\bibfield  {journal} {\bibinfo
  {journal} {Phys. Rev. B}\ }\textbf {\bibinfo {volume} {78}},\ \bibinfo
  {pages} {144404} (\bibinfo {year} {2008})}\BibitemShut {NoStop}%
\bibitem [{\citenamefont {Sato}\ \emph {et~al.}(2009)\citenamefont {Sato},
  \citenamefont {Momoi},\ and\ \citenamefont {Furusaki}}]{Sato2009}%
  \BibitemOpen
  \bibfield  {author} {\bibinfo {author} {\bibfnamefont {M.}~\bibnamefont
  {Sato}}, \bibinfo {author} {\bibfnamefont {T.}~\bibnamefont {Momoi}},\ and\
  \bibinfo {author} {\bibfnamefont {A.}~\bibnamefont {Furusaki}},\ }\bibfield
  {title} {\bibinfo {title} {{NMR relaxation rate and dynamical structure
  factors in nematic and multipolar liquids of frustrated spin chains under
  magnetic fields}},\ }\href {https://doi.org/10.1103/PhysRevB.79.060406}
  {\bibfield  {journal} {\bibinfo  {journal} {Phys. Rev. B}\ }\textbf {\bibinfo
  {volume} {79}},\ \bibinfo {pages} {060406} (\bibinfo {year}
  {2009})}\BibitemShut {NoStop}%
\bibitem [{\citenamefont {Sudan}\ \emph {et~al.}(2009)\citenamefont {Sudan},
  \citenamefont {L\"uscher},\ and\ \citenamefont {L\"auchli}}]{Sudan2009}%
  \BibitemOpen
  \bibfield  {author} {\bibinfo {author} {\bibfnamefont {J.}~\bibnamefont
  {Sudan}}, \bibinfo {author} {\bibfnamefont {A.}~\bibnamefont {L\"uscher}},\
  and\ \bibinfo {author} {\bibfnamefont {A.~M.}\ \bibnamefont {L\"auchli}},\
  }\bibfield  {title} {\bibinfo {title} {Emergent multipolar spin correlations
  in a fluctuating spiral: The frustrated ferromagnetic spin-$\frac{1}{2}$
  heisenberg chain in a magnetic field},\ }\href
  {https://doi.org/10.1103/PhysRevB.80.140402} {\bibfield  {journal} {\bibinfo
  {journal} {Phys. Rev. B}\ }\textbf {\bibinfo {volume} {80}},\ \bibinfo
  {pages} {140402} (\bibinfo {year} {2009})}\BibitemShut {NoStop}%
\bibitem [{\citenamefont {Zhitomirsky}\ and\ \citenamefont
  {Tsunetsugu}(2010)}]{Zhitomirsky2010}%
  \BibitemOpen
  \bibfield  {author} {\bibinfo {author} {\bibfnamefont {M.~E.}\ \bibnamefont
  {Zhitomirsky}}\ and\ \bibinfo {author} {\bibfnamefont {H.}~\bibnamefont
  {Tsunetsugu}},\ }\bibfield  {title} {\bibinfo {title} {Magnon pairing in
  quantum spin nematic},\ }\href {https://doi.org/10.1209/0295-5075/92/37001}
  {\bibfield  {journal} {\bibinfo  {journal} {Europhysics Letters}\ }\textbf
  {\bibinfo {volume} {92}},\ \bibinfo {pages} {37001} (\bibinfo {year}
  {2010})}\BibitemShut {NoStop}%
\bibitem [{\citenamefont {Sato}\ \emph {et~al.}(2011)\citenamefont {Sato},
  \citenamefont {Hikihara},\ and\ \citenamefont {Momoi}}]{Sato2011}%
  \BibitemOpen
  \bibfield  {author} {\bibinfo {author} {\bibfnamefont {M.}~\bibnamefont
  {Sato}}, \bibinfo {author} {\bibfnamefont {T.}~\bibnamefont {Hikihara}},\
  and\ \bibinfo {author} {\bibfnamefont {T.}~\bibnamefont {Momoi}},\ }\bibfield
   {title} {\bibinfo {title} {Field and temperature dependence of nmr
  relaxation rate in the magnetic quadrupolar liquid phase of
  spin-$\frac{1}{2}$ frustrated ferromagnetic chains},\ }\href
  {https://doi.org/10.1103/PhysRevB.83.064405} {\bibfield  {journal} {\bibinfo
  {journal} {Phys. Rev. B}\ }\textbf {\bibinfo {volume} {83}},\ \bibinfo
  {pages} {064405} (\bibinfo {year} {2011})}\BibitemShut {NoStop}%
\bibitem [{\citenamefont {Shindou}\ \emph {et~al.}(2011)\citenamefont
  {Shindou}, \citenamefont {Yunoki},\ and\ \citenamefont
  {Momoi}}]{Shindou2011}%
  \BibitemOpen
  \bibfield  {author} {\bibinfo {author} {\bibfnamefont {R.}~\bibnamefont
  {Shindou}}, \bibinfo {author} {\bibfnamefont {S.}~\bibnamefont {Yunoki}},\
  and\ \bibinfo {author} {\bibfnamefont {T.}~\bibnamefont {Momoi}},\ }\bibfield
   {title} {\bibinfo {title} {{Projective studies of spin nematics in a quantum
  frustrated ferromagnet}},\ }\href
  {https://doi.org/10.1103/PhysRevB.84.134414} {\bibfield  {journal} {\bibinfo
  {journal} {Phys. Rev. B}\ }\textbf {\bibinfo {volume} {84}},\ \bibinfo
  {pages} {134414} (\bibinfo {year} {2011})}\BibitemShut {NoStop}%
\bibitem [{\citenamefont {Momoi}\ \emph {et~al.}(2012)\citenamefont {Momoi},
  \citenamefont {Sindzingre},\ and\ \citenamefont {Kubo}}]{Momoi2012}%
  \BibitemOpen
  \bibfield  {author} {\bibinfo {author} {\bibfnamefont {T.}~\bibnamefont
  {Momoi}}, \bibinfo {author} {\bibfnamefont {P.}~\bibnamefont {Sindzingre}},\
  and\ \bibinfo {author} {\bibfnamefont {K.}~\bibnamefont {Kubo}},\ }\bibfield
  {title} {\bibinfo {title} {{Spin Nematic Order in Multiple-Spin Exchange
  Models on the Triangular Lattice}},\ }\href
  {https://doi.org/10.1103/PhysRevLett.108.057206} {\bibfield  {journal}
  {\bibinfo  {journal} {Phys. Rev. Lett.}\ }\textbf {\bibinfo {volume} {108}},\
  \bibinfo {pages} {057206} (\bibinfo {year} {2012})}\BibitemShut {NoStop}%
\bibitem [{\citenamefont {Sato}\ \emph {et~al.}(2013)\citenamefont {Sato},
  \citenamefont {Hikihara},\ and\ \citenamefont {Momoi}}]{Sato2013}%
  \BibitemOpen
  \bibfield  {author} {\bibinfo {author} {\bibfnamefont {M.}~\bibnamefont
  {Sato}}, \bibinfo {author} {\bibfnamefont {T.}~\bibnamefont {Hikihara}},\
  and\ \bibinfo {author} {\bibfnamefont {T.}~\bibnamefont {Momoi}},\ }\bibfield
   {title} {\bibinfo {title} {{Spin-Nematic and Spin-Density-Wave Orders in
  Spatially Anisotropic Frustrated Magnets in a Magnetic Field}},\ }\href
  {https://doi.org/10.1103/PhysRevLett.110.077206} {\bibfield  {journal}
  {\bibinfo  {journal} {Phys. Rev. Lett.}\ }\textbf {\bibinfo {volume} {110}},\
  \bibinfo {pages} {077206} (\bibinfo {year} {2013})}\BibitemShut {NoStop}%
\bibitem [{\citenamefont {Ueda}\ and\ \citenamefont {Momoi}(2013)}]{Ueda2013}%
  \BibitemOpen
  \bibfield  {author} {\bibinfo {author} {\bibfnamefont {H.~T.}\ \bibnamefont
  {Ueda}}\ and\ \bibinfo {author} {\bibfnamefont {T.}~\bibnamefont {Momoi}},\
  }\bibfield  {title} {\bibinfo {title} {{Nematic phase and phase separation
  near saturation field in frustrated ferromagnets}},\ }\href
  {https://doi.org/10.1103/PhysRevB.87.144417} {\bibfield  {journal} {\bibinfo
  {journal} {Phys. Rev. B}\ }\textbf {\bibinfo {volume} {87}},\ \bibinfo
  {pages} {144417} (\bibinfo {year} {2013})}\BibitemShut {NoStop}%
\bibitem [{\citenamefont {Jiang}\ \emph {et~al.}(2023)\citenamefont {Jiang},
  \citenamefont {Romh\'anyi}, \citenamefont {White}, \citenamefont
  {Zhitomirsky},\ and\ \citenamefont {Chernyshev}}]{Jiang2023}%
  \BibitemOpen
  \bibfield  {author} {\bibinfo {author} {\bibfnamefont {S.}~\bibnamefont
  {Jiang}}, \bibinfo {author} {\bibfnamefont {J.}~\bibnamefont {Romh\'anyi}},
  \bibinfo {author} {\bibfnamefont {S.~R.}\ \bibnamefont {White}}, \bibinfo
  {author} {\bibfnamefont {M.~E.}\ \bibnamefont {Zhitomirsky}},\ and\ \bibinfo
  {author} {\bibfnamefont {A.~L.}\ \bibnamefont {Chernyshev}},\ }\bibfield
  {title} {\bibinfo {title} {{Where is the Quantum Spin Nematic?}},\ }\href
  {https://doi.org/10.1103/PhysRevLett.130.116701} {\bibfield  {journal}
  {\bibinfo  {journal} {Phys. Rev. Lett.}\ }\textbf {\bibinfo {volume} {130}},\
  \bibinfo {pages} {116701} (\bibinfo {year} {2023})}\BibitemShut {NoStop}%
\bibitem [{\citenamefont {Momoi}\ and\ \citenamefont
  {Totsuka}(2000)}]{Momoi2000}%
  \BibitemOpen
  \bibfield  {author} {\bibinfo {author} {\bibfnamefont {T.}~\bibnamefont
  {Momoi}}\ and\ \bibinfo {author} {\bibfnamefont {K.}~\bibnamefont
  {Totsuka}},\ }\bibfield  {title} {\bibinfo {title} {{Magnetization plateaus
  of the Shastry-Sutherland model for
  $\mathrm{SrCu}_{2}(\mathrm{BO}_{3}{)}_{2}:$ Spin-density wave, supersolid,
  and bound states}},\ }\href {https://doi.org/10.1103/PhysRevB.62.15067}
  {\bibfield  {journal} {\bibinfo  {journal} {Phys. Rev. B}\ }\textbf {\bibinfo
  {volume} {62}},\ \bibinfo {pages} {15067} (\bibinfo {year}
  {2000})}\BibitemShut {NoStop}%
\bibitem [{\citenamefont {Wang}\ and\ \citenamefont
  {Batista}(2018)}]{Wang2018}%
  \BibitemOpen
  \bibfield  {author} {\bibinfo {author} {\bibfnamefont {Z.}~\bibnamefont
  {Wang}}\ and\ \bibinfo {author} {\bibfnamefont {C.~D.}\ \bibnamefont
  {Batista}},\ }\bibfield  {title} {\bibinfo {title} {{Dynamics and
  Instabilities of the Shastry-Sutherland Model}},\ }\href
  {https://doi.org/10.1103/PhysRevLett.120.247201} {\bibfield  {journal}
  {\bibinfo  {journal} {Phys. Rev. Lett.}\ }\textbf {\bibinfo {volume} {120}},\
  \bibinfo {pages} {247201} (\bibinfo {year} {2018})}\BibitemShut {NoStop}%
\bibitem [{\citenamefont {Yokoyama}\ and\ \citenamefont
  {Hotta}(2018)}]{Yokohama2018}%
  \BibitemOpen
  \bibfield  {author} {\bibinfo {author} {\bibfnamefont {Y.}~\bibnamefont
  {Yokoyama}}\ and\ \bibinfo {author} {\bibfnamefont {C.}~\bibnamefont
  {Hotta}},\ }\bibfield  {title} {\bibinfo {title} {Spin nematics next to spin
  singlets},\ }\href {https://doi.org/10.1103/PhysRevB.97.180404} {\bibfield
  {journal} {\bibinfo  {journal} {Phys. Rev. B}\ }\textbf {\bibinfo {volume}
  {97}},\ \bibinfo {pages} {180404} (\bibinfo {year} {2018})}\BibitemShut
  {NoStop}%
\bibitem [{\citenamefont {Hikihara}\ \emph {et~al.}(2019)\citenamefont
  {Hikihara}, \citenamefont {Misawa},\ and\ \citenamefont
  {Momoi}}]{Hikihara2019}%
  \BibitemOpen
  \bibfield  {author} {\bibinfo {author} {\bibfnamefont {T.}~\bibnamefont
  {Hikihara}}, \bibinfo {author} {\bibfnamefont {T.}~\bibnamefont {Misawa}},\
  and\ \bibinfo {author} {\bibfnamefont {T.}~\bibnamefont {Momoi}},\ }\bibfield
   {title} {\bibinfo {title} {Spin nematics in frustrated spin-dimer systems
  with bilayer structure},\ }\href
  {https://doi.org/10.1103/PhysRevB.100.214414} {\bibfield  {journal} {\bibinfo
   {journal} {Phys. Rev. B}\ }\textbf {\bibinfo {volume} {100}},\ \bibinfo
  {pages} {214414} (\bibinfo {year} {2019})}\BibitemShut {NoStop}%
\bibitem [{\citenamefont {B\"uttgen}\ \emph {et~al.}(2014)\citenamefont
  {B\"uttgen}, \citenamefont {Nawa}, \citenamefont {Fujita}, \citenamefont
  {Hagiwara}, \citenamefont {Kuhns}, \citenamefont {Prokofiev}, \citenamefont
  {Reyes}, \citenamefont {Svistov}, \citenamefont {Yoshimura},\ and\
  \citenamefont {Takigawa}}]{Buttgen2014}%
  \BibitemOpen
  \bibfield  {author} {\bibinfo {author} {\bibfnamefont {N.}~\bibnamefont
  {B\"uttgen}}, \bibinfo {author} {\bibfnamefont {K.}~\bibnamefont {Nawa}},
  \bibinfo {author} {\bibfnamefont {T.}~\bibnamefont {Fujita}}, \bibinfo
  {author} {\bibfnamefont {M.}~\bibnamefont {Hagiwara}}, \bibinfo {author}
  {\bibfnamefont {P.}~\bibnamefont {Kuhns}}, \bibinfo {author} {\bibfnamefont
  {A.}~\bibnamefont {Prokofiev}}, \bibinfo {author} {\bibfnamefont {A.~P.}\
  \bibnamefont {Reyes}}, \bibinfo {author} {\bibfnamefont {L.~E.}\ \bibnamefont
  {Svistov}}, \bibinfo {author} {\bibfnamefont {K.}~\bibnamefont {Yoshimura}},\
  and\ \bibinfo {author} {\bibfnamefont {M.}~\bibnamefont {Takigawa}},\
  }\bibfield  {title} {\bibinfo {title} {{Search for a spin-nematic phase in
  the quasi-one-dimensional frustrated magnet ${\mathrm{LiCuVO}}_{4}$}},\
  }\href {https://doi.org/10.1103/PhysRevB.90.134401} {\bibfield  {journal}
  {\bibinfo  {journal} {Phys. Rev. B}\ }\textbf {\bibinfo {volume} {90}},\
  \bibinfo {pages} {134401} (\bibinfo {year} {2014})}\BibitemShut {NoStop}%
\bibitem [{\citenamefont {Orlova}\ \emph {et~al.}(2017)\citenamefont {Orlova},
  \citenamefont {Green}, \citenamefont {Law}, \citenamefont {Gorbunov},
  \citenamefont {Chanda}, \citenamefont {Kr\"amer}, \citenamefont
  {Horvati\ifmmode~\acute{c}\else \'{c}\fi{}}, \citenamefont {Kremer},
  \citenamefont {Wosnitza},\ and\ \citenamefont {Rikken}}]{Orlova2017}%
  \BibitemOpen
  \bibfield  {author} {\bibinfo {author} {\bibfnamefont {A.}~\bibnamefont
  {Orlova}}, \bibinfo {author} {\bibfnamefont {E.~L.}\ \bibnamefont {Green}},
  \bibinfo {author} {\bibfnamefont {J.~M.}\ \bibnamefont {Law}}, \bibinfo
  {author} {\bibfnamefont {D.~I.}\ \bibnamefont {Gorbunov}}, \bibinfo {author}
  {\bibfnamefont {G.}~\bibnamefont {Chanda}}, \bibinfo {author} {\bibfnamefont
  {S.}~\bibnamefont {Kr\"amer}}, \bibinfo {author} {\bibfnamefont
  {M.}~\bibnamefont {Horvati\ifmmode~\acute{c}\else \'{c}\fi{}}}, \bibinfo
  {author} {\bibfnamefont {R.~K.}\ \bibnamefont {Kremer}}, \bibinfo {author}
  {\bibfnamefont {J.}~\bibnamefont {Wosnitza}},\ and\ \bibinfo {author}
  {\bibfnamefont {G.~L. J.~A.}\ \bibnamefont {Rikken}},\ }\bibfield  {title}
  {\bibinfo {title} {{Nuclear Magnetic Resonance Signature of the Spin-Nematic
  Phase in ${\mathrm{LiCuVO}}_{4}$ at High Magnetic Fields}},\ }\href
  {https://doi.org/10.1103/PhysRevLett.118.247201} {\bibfield  {journal}
  {\bibinfo  {journal} {Phys. Rev. Lett.}\ }\textbf {\bibinfo {volume} {118}},\
  \bibinfo {pages} {247201} (\bibinfo {year} {2017})}\BibitemShut {NoStop}%
\bibitem [{\citenamefont {Yoshida}\ \emph {et~al.}(2017)\citenamefont
  {Yoshida}, \citenamefont {Nawa}, \citenamefont {Ishikawa}, \citenamefont
  {Takigawa}, \citenamefont {Jeong}, \citenamefont {Kr\"amer}, \citenamefont
  {Horvati\ifmmode~\acute{c}\else \'{c}\fi{}}, \citenamefont {Berthier},
  \citenamefont {Matsui}, \citenamefont {Goto}, \citenamefont {Kimura},
  \citenamefont {Sasaki}, \citenamefont {Yamaura}, \citenamefont {Yoshida},
  \citenamefont {Okamoto},\ and\ \citenamefont {Hiroi}}]{Yoshida2017}%
  \BibitemOpen
  \bibfield  {author} {\bibinfo {author} {\bibfnamefont {M.}~\bibnamefont
  {Yoshida}}, \bibinfo {author} {\bibfnamefont {K.}~\bibnamefont {Nawa}},
  \bibinfo {author} {\bibfnamefont {H.}~\bibnamefont {Ishikawa}}, \bibinfo
  {author} {\bibfnamefont {M.}~\bibnamefont {Takigawa}}, \bibinfo {author}
  {\bibfnamefont {M.}~\bibnamefont {Jeong}}, \bibinfo {author} {\bibfnamefont
  {S.}~\bibnamefont {Kr\"amer}}, \bibinfo {author} {\bibfnamefont
  {M.}~\bibnamefont {Horvati\ifmmode~\acute{c}\else \'{c}\fi{}}}, \bibinfo
  {author} {\bibfnamefont {C.}~\bibnamefont {Berthier}}, \bibinfo {author}
  {\bibfnamefont {K.}~\bibnamefont {Matsui}}, \bibinfo {author} {\bibfnamefont
  {T.}~\bibnamefont {Goto}}, \bibinfo {author} {\bibfnamefont {S.}~\bibnamefont
  {Kimura}}, \bibinfo {author} {\bibfnamefont {T.}~\bibnamefont {Sasaki}},
  \bibinfo {author} {\bibfnamefont {J.}~\bibnamefont {Yamaura}}, \bibinfo
  {author} {\bibfnamefont {H.}~\bibnamefont {Yoshida}}, \bibinfo {author}
  {\bibfnamefont {Y.}~\bibnamefont {Okamoto}},\ and\ \bibinfo {author}
  {\bibfnamefont {Z.}~\bibnamefont {Hiroi}},\ }\bibfield  {title} {\bibinfo
  {title} {{Spin dynamics in the high-field phases of volborthite}},\ }\href
  {https://doi.org/10.1103/PhysRevB.96.180413} {\bibfield  {journal} {\bibinfo
  {journal} {Phys. Rev. B}\ }\textbf {\bibinfo {volume} {96}},\ \bibinfo
  {pages} {180413} (\bibinfo {year} {2017})}\BibitemShut {NoStop}%
\bibitem [{\citenamefont {Nath}\ \emph {et~al.}(2008)\citenamefont {Nath},
  \citenamefont {Tsirlin}, \citenamefont {Rosner},\ and\ \citenamefont
  {Geibel}}]{Nath2008}%
  \BibitemOpen
  \bibfield  {author} {\bibinfo {author} {\bibfnamefont {R.}~\bibnamefont
  {Nath}}, \bibinfo {author} {\bibfnamefont {A.~A.}\ \bibnamefont {Tsirlin}},
  \bibinfo {author} {\bibfnamefont {H.}~\bibnamefont {Rosner}},\ and\ \bibinfo
  {author} {\bibfnamefont {C.}~\bibnamefont {Geibel}},\ }\bibfield  {title}
  {\bibinfo {title} {{Magnetic properties of
  $\text{BaCdVO}{({\text{PO}}_{4})}_{2}$: A strongly frustrated spin-$1/2$
  square lattice close to the quantum critical regime}},\ }\href
  {https://doi.org/10.1103/PhysRevB.78.064422} {\bibfield  {journal} {\bibinfo
  {journal} {Phys. Rev. B}\ }\textbf {\bibinfo {volume} {78}},\ \bibinfo
  {pages} {064422} (\bibinfo {year} {2008})}\BibitemShut {NoStop}%
\bibitem [{\citenamefont {Skoulatos}\ \emph {et~al.}(2019)\citenamefont
  {Skoulatos}, \citenamefont {Rucker}, \citenamefont {Nilsen}, \citenamefont
  {Bertin}, \citenamefont {Pomjakushina}, \citenamefont {Ollivier},
  \citenamefont {Schneidewind}, \citenamefont {Georgii}, \citenamefont
  {Zaharko}, \citenamefont {Keller}, \citenamefont {R\"uegg}, \citenamefont
  {Pfleiderer}, \citenamefont {Schmidt}, \citenamefont {Shannon}, \citenamefont
  {Kriele}, \citenamefont {Senyshyn},\ and\ \citenamefont
  {Smerald}}]{Skoulatos2019}%
  \BibitemOpen
  \bibfield  {author} {\bibinfo {author} {\bibfnamefont {M.}~\bibnamefont
  {Skoulatos}}, \bibinfo {author} {\bibfnamefont {F.}~\bibnamefont {Rucker}},
  \bibinfo {author} {\bibfnamefont {G.~J.}\ \bibnamefont {Nilsen}}, \bibinfo
  {author} {\bibfnamefont {A.}~\bibnamefont {Bertin}}, \bibinfo {author}
  {\bibfnamefont {E.}~\bibnamefont {Pomjakushina}}, \bibinfo {author}
  {\bibfnamefont {J.}~\bibnamefont {Ollivier}}, \bibinfo {author}
  {\bibfnamefont {A.}~\bibnamefont {Schneidewind}}, \bibinfo {author}
  {\bibfnamefont {R.}~\bibnamefont {Georgii}}, \bibinfo {author} {\bibfnamefont
  {O.}~\bibnamefont {Zaharko}}, \bibinfo {author} {\bibfnamefont
  {L.}~\bibnamefont {Keller}}, \bibinfo {author} {\bibfnamefont
  {C.}~\bibnamefont {R\"uegg}}, \bibinfo {author} {\bibfnamefont
  {C.}~\bibnamefont {Pfleiderer}}, \bibinfo {author} {\bibfnamefont
  {B.}~\bibnamefont {Schmidt}}, \bibinfo {author} {\bibfnamefont
  {N.}~\bibnamefont {Shannon}}, \bibinfo {author} {\bibfnamefont
  {A.}~\bibnamefont {Kriele}}, \bibinfo {author} {\bibfnamefont
  {A.}~\bibnamefont {Senyshyn}},\ and\ \bibinfo {author} {\bibfnamefont
  {A.}~\bibnamefont {Smerald}},\ }\bibfield  {title} {\bibinfo {title}
  {{Putative spin-nematic phase in
  $\mathrm{BaCdVO}({\mathrm{PO}}_{4}{)}_{2}$}},\ }\href
  {https://doi.org/10.1103/PhysRevB.100.014405} {\bibfield  {journal} {\bibinfo
   {journal} {Phys. Rev. B}\ }\textbf {\bibinfo {volume} {100}},\ \bibinfo
  {pages} {014405} (\bibinfo {year} {2019})}\BibitemShut {NoStop}%
\bibitem [{\citenamefont {Povarov}\ \emph {et~al.}(2019)\citenamefont
  {Povarov}, \citenamefont {Bhartiya}, \citenamefont {Yan},\ and\ \citenamefont
  {Zheludev}}]{Povarov2019}%
  \BibitemOpen
  \bibfield  {author} {\bibinfo {author} {\bibfnamefont {K.~Y.}\ \bibnamefont
  {Povarov}}, \bibinfo {author} {\bibfnamefont {V.~K.}\ \bibnamefont
  {Bhartiya}}, \bibinfo {author} {\bibfnamefont {Z.}~\bibnamefont {Yan}},\ and\
  \bibinfo {author} {\bibfnamefont {A.}~\bibnamefont {Zheludev}},\ }\bibfield
  {title} {\bibinfo {title} {Thermodynamics of a frustrated quantum magnet on a
  square lattice},\ }\href {https://doi.org/10.1103/PhysRevB.99.024413}
  {\bibfield  {journal} {\bibinfo  {journal} {Phys. Rev. B}\ }\textbf {\bibinfo
  {volume} {99}},\ \bibinfo {pages} {024413} (\bibinfo {year}
  {2019})}\BibitemShut {NoStop}%
\bibitem [{\citenamefont {Momoi}(2024)}]{Momoi2024}%
  \BibitemOpen
  \bibfield  {author} {\bibinfo {author} {\bibfnamefont {T.}~\bibnamefont
  {Momoi}},\ }\bibfield  {title} {\bibinfo {title} {Dynamics of quantum
  spin-nematics: Comparisons with canted antiferromagnets},\ }\href
  {https://doi.org/10.1103/PhysRevResearch.6.013169} {\bibfield  {journal}
  {\bibinfo  {journal} {Phys. Rev. Res.}\ }\textbf {\bibinfo {volume} {6}},\
  \bibinfo {pages} {013169} (\bibinfo {year} {2024})}\BibitemShut {NoStop}%
\bibitem [{\citenamefont {Hirobe}\ \emph {et~al.}(2019)\citenamefont {Hirobe},
  \citenamefont {Sato}, \citenamefont {Hagihala}, \citenamefont {Shiomi},
  \citenamefont {Masuda},\ and\ \citenamefont {Saitoh}}]{Hirobe2019}%
  \BibitemOpen
  \bibfield  {author} {\bibinfo {author} {\bibfnamefont {D.}~\bibnamefont
  {Hirobe}}, \bibinfo {author} {\bibfnamefont {M.}~\bibnamefont {Sato}},
  \bibinfo {author} {\bibfnamefont {M.}~\bibnamefont {Hagihala}}, \bibinfo
  {author} {\bibfnamefont {Y.}~\bibnamefont {Shiomi}}, \bibinfo {author}
  {\bibfnamefont {T.}~\bibnamefont {Masuda}},\ and\ \bibinfo {author}
  {\bibfnamefont {E.}~\bibnamefont {Saitoh}},\ }\bibfield  {title} {\bibinfo
  {title} {{Magnon Pairs and Spin-Nematic Correlation in the Spin Seebeck
  Effect}},\ }\href {https://doi.org/10.1103/PhysRevLett.123.117202} {\bibfield
   {journal} {\bibinfo  {journal} {Phys. Rev. Lett.}\ }\textbf {\bibinfo
  {volume} {123}},\ \bibinfo {pages} {117202} (\bibinfo {year}
  {2019})}\BibitemShut {NoStop}%
\bibitem [{\citenamefont {Lacroix}\ \emph {et~al.}(2011)\citenamefont
  {Lacroix}, \citenamefont {Mendels},\ and\ \citenamefont
  {Mila}}]{Lacroix2011}%
  \BibitemOpen
  \bibfield  {author} {\bibinfo {author} {\bibfnamefont {C.}~\bibnamefont
  {Lacroix}}, \bibinfo {author} {\bibfnamefont {P.}~\bibnamefont {Mendels}},\
  and\ \bibinfo {author} {\bibfnamefont {F.}~\bibnamefont {Mila}},\ }\href@noop
  {} {\emph {\bibinfo {title} {{Introduction to Frustrated Magnetism:
  Materials, Experiments, Theory}}}},\ Springer Series in Solid-State Sciences\
  (\bibinfo  {publisher} {Springer Berlin Heidelberg},\ \bibinfo {year}
  {2011})\BibitemShut {NoStop}%
\bibitem [{\citenamefont {Kato}\ \emph {et~al.}(2019)\citenamefont {Kato},
  \citenamefont {Ohnuma}, \citenamefont {Matsuo}, \citenamefont {Rech},
  \citenamefont {Jonckheere},\ and\ \citenamefont {Martin}}]{Kato2019}%
  \BibitemOpen
  \bibfield  {author} {\bibinfo {author} {\bibfnamefont {T.}~\bibnamefont
  {Kato}}, \bibinfo {author} {\bibfnamefont {Y.}~\bibnamefont {Ohnuma}},
  \bibinfo {author} {\bibfnamefont {M.}~\bibnamefont {Matsuo}}, \bibinfo
  {author} {\bibfnamefont {J.}~\bibnamefont {Rech}}, \bibinfo {author}
  {\bibfnamefont {T.}~\bibnamefont {Jonckheere}},\ and\ \bibinfo {author}
  {\bibfnamefont {T.}~\bibnamefont {Martin}},\ }\bibfield  {title} {\bibinfo
  {title} {Microscopic theory of spin transport at the interface between a
  superconductor and a ferromagnetic insulator},\ }\href
  {https://doi.org/10.1103/PhysRevB.99.144411} {\bibfield  {journal} {\bibinfo
  {journal} {Phys. Rev. B}\ }\textbf {\bibinfo {volume} {99}},\ \bibinfo
  {pages} {144411} (\bibinfo {year} {2019})}\BibitemShut {NoStop}%
\bibitem [{\citenamefont {Seyed~Heydari}\ \emph {et~al.}(2025)\citenamefont
  {Seyed~Heydari}, \citenamefont {Belzig},\ and\ \citenamefont
  {Rohling}}]{Heydari2025}%
  \BibitemOpen
  \bibfield  {author} {\bibinfo {author} {\bibfnamefont {M.}~\bibnamefont
  {Seyed~Heydari}}, \bibinfo {author} {\bibfnamefont {W.}~\bibnamefont
  {Belzig}},\ and\ \bibinfo {author} {\bibfnamefont {N.}~\bibnamefont
  {Rohling}},\ }\bibfield  {title} {\bibinfo {title} {Influence of disorder at
  insulator-metal interface on spin transport},\ }\href
  {https://doi.org/10.1103/PhysRevB.111.054414} {\bibfield  {journal} {\bibinfo
   {journal} {Phys. Rev. B}\ }\textbf {\bibinfo {volume} {111}},\ \bibinfo
  {pages} {054414} (\bibinfo {year} {2025})}\BibitemShut {NoStop}%
\bibitem [{\citenamefont {Yama}\ \emph {et~al.}(2025)\citenamefont {Yama},
  \citenamefont {Matsuo},\ and\ \citenamefont {Kato}}]{Yama2025}%
  \BibitemOpen
  \bibfield  {author} {\bibinfo {author} {\bibfnamefont {M.}~\bibnamefont
  {Yama}}, \bibinfo {author} {\bibfnamefont {M.}~\bibnamefont {Matsuo}},\ and\
  \bibinfo {author} {\bibfnamefont {T.}~\bibnamefont {Kato}},\ }\bibfield
  {title} {\bibinfo {title} {Microscopic theory of rashba-edelstein
  magnetoresistance},\ }\href {https://doi.org/10.1103/PhysRevB.111.144416}
  {\bibfield  {journal} {\bibinfo  {journal} {Phys. Rev. B}\ }\textbf {\bibinfo
  {volume} {111}},\ \bibinfo {pages} {144416} (\bibinfo {year}
  {2025})}\BibitemShut {NoStop}%
\bibitem [{\citenamefont {Mizukami}\ \emph {et~al.}(2016)\citenamefont
  {Mizukami}, \citenamefont {Sugihara}, \citenamefont {Iihama}, \citenamefont
  {Sasaki}, \citenamefont {Suzuki},\ and\ \citenamefont
  {Miyazaki}}]{Mizukami2016}%
  \BibitemOpen
  \bibfield  {author} {\bibinfo {author} {\bibfnamefont {S.}~\bibnamefont
  {Mizukami}}, \bibinfo {author} {\bibfnamefont {A.}~\bibnamefont {Sugihara}},
  \bibinfo {author} {\bibfnamefont {S.}~\bibnamefont {Iihama}}, \bibinfo
  {author} {\bibfnamefont {Y.}~\bibnamefont {Sasaki}}, \bibinfo {author}
  {\bibfnamefont {K.}~\bibnamefont {Suzuki}},\ and\ \bibinfo {author}
  {\bibfnamefont {T.}~\bibnamefont {Miyazaki}},\ }\bibfield  {title} {\bibinfo
  {title} {{Laser-induced THz magnetization precession for a tetragonal
  Heusler-like nearly compensated ferrimagnet}},\ }\bibfield  {journal}
  {\bibinfo  {journal} {Appl. Phys. Lett.}\ }\textbf {\bibinfo {volume}
  {108}},\ \href {https://doi.org/10.1063/1.4939447} {10.1063/1.4939447}
  (\bibinfo {year} {2016})\BibitemShut {NoStop}%
\bibitem [{\citenamefont {Walowski}\ and\ \citenamefont
  {Münzenberg}(2016)}]{Walowski2016}%
  \BibitemOpen
  \bibfield  {author} {\bibinfo {author} {\bibfnamefont {J.}~\bibnamefont
  {Walowski}}\ and\ \bibinfo {author} {\bibfnamefont {M.}~\bibnamefont
  {Münzenberg}},\ }\bibfield  {title} {\bibinfo {title} {Perspective:
  Ultrafast magnetism and thz spintronics},\ }\href
  {https://doi.org/10.1063/1.4958846} {\bibfield  {journal} {\bibinfo
  {journal} {J. Appl. Phys.}\ }\textbf {\bibinfo {volume} {120}},\ \bibinfo
  {pages} {140901} (\bibinfo {year} {2016})}\BibitemShut {NoStop}%
\end{thebibliography}%
\clearpage

\end{document}